\documentclass[a4paper]{article}
\usepackage{a4wide}
\usepackage{graphicx}
\usepackage[dvipsnames]{xcolor}
\graphicspath{{./images/}}
\usepackage{caption}
\usepackage{empheq}
\usepackage{amsthm}
\usepackage{amssymb, latexsym, mathrsfs}
\usepackage{amsmath}
\usepackage{dsfont}
\usepackage{enumerate}
\usepackage{algorithm}
\usepackage{color}
\usepackage{transparent}
\usepackage{subfig}
\usepackage{url}
\usepackage[affil-it]{authblk}
\usepackage{booktabs}
\usepackage{multirow}
\usepackage{float} 
\usepackage[showframe=false]{geometry}
\usepackage{enumitem}

\usepackage{graphicx}
\usepackage{epsfig}
\usepackage{pgfplots}
\usepackage{tikz}
\newlength\fheight 
\newlength\fwidth 
\usepackage{indentfirst}
\pgfplotsset{compat=newest} 
\pgfplotsset{plot coordinates/math parser=false}
\usepackage{tikzscale}
\usepackage[american resistor, european voltage, european current]{circuitikz}
\usetikzlibrary{arrows,shapes,positioning}
\usetikzlibrary{decorations.markings,decorations.pathmorphing,
	decorations.pathreplacing}
\usetikzlibrary{calc,patterns,shapes.geometric}
\usetikzlibrary{arrows}
\usetikzlibrary{shapes.misc}
\usetikzlibrary{arrows.meta,positioning}

\usepackage{titlesec}
\titleformat*{\section}{\Large\bfseries}
\titleformat*{\subsection}{\large \bfseries}
\titleformat*{\subsubsection}{\Small\bfseries}
\titleformat*{\paragraph}{\Small\bfseries}
\titleformat*{\subparagraph}{\Small\bfseries}
\date{}

\usepackage{amsthm}
\usepackage{amssymb}
\theoremstyle{plain}
\newtheorem{theorem}{Theorem}[section]

\newtheorem{lemma}{Lemma}[section]
\newtheorem{assum}{Assumption}[section]
\newtheorem{remark}{Remark}[section]
\newtheorem {proposition}{Proposition}[section]	

\theoremstyle{definition}

\newcommand\blfootnote[1]{%
	\begingroup
	\renewcommand\thefootnote{}\footnote{#1}%
	\addtocounter{footnote}{-1}%
	\endgroup
}

\newcommand{\BB}{{\mathcal{B}}}

\newcommand{\DD}{{\mathcal{D}}}

\newcommand{\GG}{{\mathcal{G}}}

\newcommand{\JJ}{{\mathcal{J}}}

\newcommand{\LL}{{\mathcal{L}}}

\newcommand{\PP}{{\mathcal{P}}}

\newcommand{\VV}{{\mathcal{V}}}

\newcommand{\XX}{{\mathcal{X}}}
\newcommand{\YY}{{\mathcal{Y}}}

\newcommand{\mbf}[1]{\mathbf{#1}}                  

 \newcommand{\ALBREV}[1]{{\color{black} #1}}
 \newcommand{\PNREV}[1]{{\color{black} #1}}
 
\begin{document}
	\title{\LARGE \bf Hierarchical Control in Islanded DC Microgrids with Flexible Structures}

          \author[1]{Pulkit Nahata$^{* ,}$}
           \author[2]{Alessio La Bella$^{* ,}$}
           \author[2]{Riccardo Scattolini}
     \author[1]{Giancarlo Ferrari-Trecate}

     \affil[1]{\small Automatic Control Laboratory, \'Ecole Polytechnique F\'ed\'erale de Lausanne, Lausanne, Switzerland}
          \affil[2]{\small Dipartimento di Elettronica, Informazione e Bioingegneria, Politecnico di Milano, Italy}

     \date{\textbf{Technical Report}\\ September, 2019\blfootnote{indicates equal contribution. This work has received support from  the Swiss National Science Foundation under the COFLEX project (grant number 200021{\_}169906) and the Research Fund for the Italian Electrical System in compliance with the Decree of Minister of Economic Development April 16, 2018. Electronic addresses: \texttt{\{pulkit.nahata,~giancarlo.ferraritrecate\}@epfl.ch}, \texttt{\{alessio.labella,riccardo.scattolini\}@polimi.it}}}

     \maketitle
     \begin{abstract}
Hierarchical architectures stacking primary, secondary, and tertiary layers are widely employed for the operation and control of islanded DC microgrids (DCmGs), composed of Distribution Generation Units (DGUs), loads, and power lines. However, a comprehensive analysis of all the layers put together is often missing. In this work, we remedy this limitation by setting out a top-to-bottom hierarchical control architecture. Decentralized voltage controllers attached to DGUs form our primary layer. Governed by an MPC--based Energy Management System (EMS), our tertiary layer generates optimal power references and decision variables for DGUs. In particular, decision variables can turn DGUs ON/OFF and select their operation modes. An intermediary secondary layer translates EMS power references into appropriate voltage signals required by the primary layer. More specifically, to provide a voltage solution, the secondary layer solves an optimization problem embedding power-flow equations shown to be always solvable.  Since load voltages are not directly enforced, their uniqueness is necessary for DGUs to produce reference powers handed down by the EMS. To this aim, we deduce a novel uniqueness condition based only on local load parameters. Our control framework, besides being applicable for generic DCmG topologies, can accommodate topological changes caused by EMS commands. Its functioning is validated via simulations on a modified 16-bus DC system.
     \end{abstract}

\newpage
\section{Introduction}
\PNREV{Microgrids (mGs) are small-scale electric networks consisting of Distributed Generation Units (DGUs) and different loads. Apart from their manifold advantages like integration of renewables, enhanced power quality, reduced transmission losses and capability to operate in grid-connected and islanded modes, mGs are also compatible with both AC and  DC operating standard \cite{la2018two,Shafiee2014,babazadeh2013robust}.} In particular, DC microgrids (DCmGs), have gained traction in recent times. Their rising popularity can be attributed to the development of efficient converters, natural interface with many renewable energy sources (for instance PV modules), batteries, and many electronic loads (various appliances, LEDs, electric vehicles, computers etc), inherently DC in nature \cite{Dragicevic1, Meng}. 

\PNREV{A stable and economic operation of an islanded DC microgrid (DCmG) is a multi-objective problem. As such, it necessitates a control entity which properly regulates the internal voltages and efficiently coordinates DGU operations while taking into consideration the non-deterministic absorption/production of loads and renewables. To this aim, a hierarchical architecture spanning different control stages, time scales, and physical layers is often employed \cite{Dragicevic1,Meng,Bidram,la2017hierarchical}.}

Generally, a primary control layer, acting at the component level, is responsible for voltage stability, crucial in islanded DCmGs interfaced with nonlinear loads \cite{Nahata}. Many research studies have aimed at designing decentralized stabilizing primary controllers, implemented at each DGU with a view to tracking constant voltage references in steady state. \PNREV{For this purpose, different techniques such as droop control \cite{Shafiee2014, Dragicevic1}, plug-and-play \cite{Martinelli2018, Nahata, strehle2020scalable} and sliding-mode control \cite{cucuzzella2017decentralized} have been explored. Being blind voltage emulators, primary controllers are incapable of incorporating various operational and economic constraints needed to sustain a continuous and proper functioning of an islanded DCmG. High-level supervisory control architectures are, therefore, necessary to coordinate primary voltage references. Consensus-based supervisory controllers discussed in \cite{Tucci2018, NahataECC} appropriately tweak primary voltage references to attain proportional load sharing and voltage balancing. Despite presenting the major advantage of having a distributed structure, these controllers assume load satisfiability and unsaturated inputs at all times. Supervising control strategies considering saturated inputs for primary voltage controllers are proposed in \cite{martinelli2019secondary}; however, capability limits, operation mode, and filter losses of DGUs, along with optimal power dispatch are neglected. One can overcome the aforementioned limitations by designing an Energy Management System (EMS), which can meet specified power and energy management strategies while respecting generation constraints and other economic objectives like optimal power dispatch, load sharing, and battery management. Flowchart-based \PNREV{EMSs} encompassing multiple case scenarios are discussed in \cite{Kumar, Dragicevic3}, whereas the use of optimization methods and predictive algorithms to design an EMS is investigated  in \cite{la2017hierarchical, marzband2017optimal}.} 

\PNREV{In general, EMSs utilize power balance equations to provide optimal power set-points to the {DGUs}, especially when based on complex optimization algorithms such as stochastic or mixed-integer techniques as in \cite{Bemporad, parisio2016stochastic, cominesi2016multi, Iovine2019, hans2019risk, michaelson2017predictive}.} Nevertheless, when the primary layer is voltage controlled, these \PNREV{EMSs} cannot be directly implemented in \PNREV{DCmGs} as the optimal power references must somehow be translated into suitable voltage set-points.  Such a translation is not straightforward for mGs with meshed topologies and, effectively, requires the solution of power-flow equations. Moreover, considering that the voltages can solely be enforced by the DGUs, a unique voltage equilibrium may fail to exist at the load buses in the presence of nonlinear loads (e.g. constant power loads) \cite{matveev2018existence}.

\subsection{Main Contributions}
\begin{figure}[!t]
	\centering
	\begin{footnotesize}
		\begin{tikzpicture}[scale=0.85]
		
		\node (E)at (0,2) [draw, align=center, minimum width=5cm]{Tertiary Control \\ (EMS)};
		
		\node (S)at (0,0) [draw, align=center, minimum width=5cm]{Secondary Control };
		
		\node (C1)at (-2.5,-2) [draw, align=center,minimum width=1cm]{$\mathcal{C}_{[i]}$};
		\node (C2)at (2.5,-2) [draw, align=center, minimum width=1cm]{$\mathcal{C}_{[j]}$};
		\node (DGU1)at (-2.5,-3) [draw, align=center, minimum width=1.5cm]{DGU $i$};
		\node (DGU2)at (2.5,-3) [draw, align=center, minimum width=1.5cm]{DGU $j$};
		
		\draw[-latex,thick] ($(E.south)-(1,0)$) --node[anchor=east] {$\bar{P}_{G,i},\delta_i$} ($(S.north)-(1,0)$) ;
		\draw (0,1) node {$\cdots$};
		\draw (0,-0.625) node {$\cdots$};
		\draw[-latex,thick] ($(E.south)+(1,0)$) --node[anchor=west] {$\bar{P}_{G,j},\delta_j$} ($(S.north)+(1,0)$);
		\draw[-latex,thick] ($(S.south)-(1,0)$) --node[ sloped, anchor=south, above] {$V_{i}^*$} (C1) ;
		\draw[-latex,thick] ($(S.south)+(1,0)$) --node[ sloped, anchor=south, above] {$V_{j}^*$} (C2) ;
		\draw[-latex,thick] (C1) -- (DGU1) ;
		\draw[-latex,thick] (C2) -- (DGU2) ;
		\draw[-latex,thick] (DGU1) -- (-2.5,-4.3) ;
		\draw[-latex,thick] (DGU2) -- (2.5,-4.32) ;
		\draw[-latex,thick] (DGU1.east) -- (-1,-3) -- (-1,-2) -- (C1.east);
		\draw[-latex,thick] (DGU2.west) -- (1,-3) -- (1,-2) -- (C2.west);
		\node[cloud, cloud puffs=16.2, cloud ignores aspect, minimum width=5.26cm, minimum height=1.5cm, opacity=0.75, align=center, fill=gray!40,draw] (Gcloud) at (0, -5) {DC Microgrid  Network \\ ZIP Loads};
		\draw[-latex,thick]  (4.25,-5)-- node[sloped, midway,below]{$\bar{I}_L,\,\bar{P}_L$} (4.25,0) -- (S.east) ; 
		\draw[-latex,thick] (4.25,0) -- (4.25,1.75) -- ($(E.east)-(0,0.25)$) ;
		\draw[-latex,thick] (3.27,-5) -- (5,-5)-- node[sloped, midway,below]{{${S}_B$,\,${P}^o_{PV}$}} (5,0) -- (5,2.25) -- ($(E.east)+(0,0.25)$) ;
		\draw [-latex,thick] (0,3.25) node[anchor=north, above]{$P_{PV}^{f}$} -- (E.north);
		\draw [-latex,thick] (-0.75,3.25) node[anchor=north, above]{$\bar{P}^f_{L}$} -- ($(E.north)-(0.75,0)$);
		\draw [-latex,thick] (0.75,3.25) node[anchor=north, above]{$\bar{I}^f_L$} -- ($(E.north)+(0.75,0)$);
		\draw[black, dashed] (-3.75,-1.1)  -- node[midway,below]{Primary Control} (3.75,.-1.1) -- (3.75,-3.8) -- (-3.75,-3.8)   -- (-3.75,-1.1); 
		\end{tikzpicture} 
	\end{footnotesize}
	\caption{Hierarchical control scheme for DC microgrids.}
	\vspace*{-3mm}
	\label{fig:hierarchy}
\end{figure}
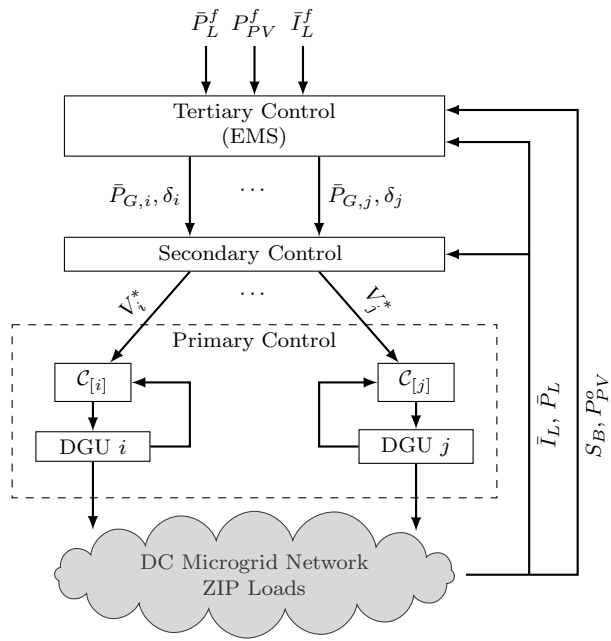

\PNREV{The lack of a detailed, all-round control scheme defining the interface between, and exact roles of, multiple control layers  motivated the design of a comprehensive three-layered hierarchical control architecture for the overall operation and control of an islanded DCmG with flexible topologies. A schematic of the proposed architecture is depicted in Figure \ref{fig:hierarchy}; a detailed description of the variables in Figure \ref{fig:hierarchy} is presented in Sections \ref{sec:EMS} and \ref{sec:secondary}.

\begin{itemize}
	\item \textit{{Tertiary layer}}:  An MPC-based EMS sits at the uppermost level, and defines power references and operation modes for DGUs. The objective of this layer is to minimize cost of production and to optimize DGU usage, e.g. avoiding frequent changes in operation mode. The EMS accounts for DGU saturation limits, type-based DGU power models, and load/generation forecasts. To efficiently solve the MPC problem, this layer discards network and DGU filter losses---nonlinear and non-convex.

	\smallskip
	\item \textit{{Secondary layer}}: A novel secondary control scheme, acting as an interface between the primary and tertiary layers, is designed to translate EMS power references into apposite voltage references. As DGUs are controlled by primary voltage regulators, such an operation is necessary for the EMS power references to be effectively produced/absorbed in the DCmG network. This task is performed via a static optimization problem subject to the network DCmG power-flow equations embedding network and DGU filter losses, as well as DGU capability limits and DCmG voltage constraints. The secondary layer is executed at a faster time scale with respect to the EMS, ensuring quick restoration of the EMS power references in case of changes in load absorption.
	
	\smallskip
	\item \textit{{Primary layer}}: Each DGU is provided with a decentralized primary voltage controller which, along with other such controllers, makes the primary layer. \\
\end{itemize}
}
In order to leverage the advances in grid-stabilizing decentralized primary voltage control, we assume that all DGUs are equipped with primary voltage regulators. The structure and design of primary voltage controllers along with stability certificates and proofs are skipped in this work.  A detailed analysis can be found in \cite{Martinelli2018, Nahata, strehle2020scalable, cucuzzella2017decentralized} which show control design based on the plug-and-play paradigm allowing DGUs to effortlessly enter/leave the DCmG without spoiling overall voltage stability. 

\PNREV{The proposed hierarchical control system is applied to a DCmG with DGUs interfaced with nonrenewable dispatchable resources, batteries, and PV modules}. The  EMS at the tertiary layer generates optimal power references and decides the operation modes of DGUs by solving an MPC mixed-integer problem at every sampling instant while taking into account forecasts and system parameters. The generated decision variables  variables serve to turn ON/OFF dispatchable  DGUs, switch the PV DGUs between Maximum Power Point Tracking (MPPT) and power curtailment modes, and control the charging/discharging of battery DGUs. In spite of a change in topology that may take place due to EMS commands, the collective voltage stability of the DCmG network is ensured by decentralized plug-and-play primary controllers.
 
\PNREV{To perform a power-voltage translation, the secondary controller makes use of an optimization problem subject to DCmG power-flow equations. Although these equations---indispensable to the feasibility of the secondary optimization problem---are nonlinear and non-convex, we prove that a solution to these can be always computed. We point out that the existence of a solution to the power-flow equations has also been addressed in \cite{simpson2016voltage, taheri2018power} Nevertheless, the tools therein cannot be used directly as, in our case, the DGU voltage references are free optimization variables and not known \textit{a priori}. Furthermore, as a complement, we also state a necessary condition for the solvability of the stated optimization problem.}

We highlight that the voltages can be enforced only at DGU nodes and therefore, the uniqueness of voltages appearing at the load nodes is necessary for the attainment of predefined operational objectives. Indeed, if the load voltages are different from those anticipated by the secondary layer, permissible voltage limits may be violated and DGUs may fail to generate the optimal power set-points provided by the EMS. In this respect, we provide a novel condition for the uniqueness of load voltages and DGU power injections. The uniqueness of voltages has also been addressed in \cite{simpson2016voltage}, where the deduced condition depends on the generator voltages and the topological parameters of the overall network. Here, we provide a novel and simpler condition which relies only on local load parameters and can easily be verified.

\PNREV{Different from the hierarchical control approaches for islanded DCmG presented in \cite{Shafiee2014,Kumar, Dragicevic3, Iovine2019, han2019review, jin2013implementation, ma2016transmission,babazadeh2019distributed}, our scheme works for any DCmG topology, considers nonlinear ZIP (constant impedance, constant current, and constant power) loads, and accommodates temporal topological variations caused by the addition/removal of certain DGUs. Furthermore, none of the foregoing contributions simultaneously ensure optimal management of DGUs (MPC-based EMS) and stable regulation of voltages (decentralized primary controllers). For instance, the hierarchical control methodologies presented in \cite{jin2013implementation, ma2016transmission,babazadeh2019distributed} address voltage regulation and optimization of power flows, but discard energy management and type-based DGU power models. On the other hand, \mbox{MPC-based EMSs} are proposed in \cite{Bemporad,hans2019risk,michaelson2017predictive}; but voltage stability is not addressed.}

Preliminary results of this work have been reported in \cite{LaBella} where (i) design of an EMS was deferred to future work, (ii) interface between secondary and tertiary layers was not discussed, \PNREV{(iii)} no distinction was made on the type of DGU; DGU dynamics were not modeled, and (iv) detailed proofs of main theorems and propositions were skipped. Furthermore, this article demonstrates a coordinated operation of multiple control layers on a 16-node DCmG. 

The structure of DCmG along with proposed hierarchical control scheme is described in {Section} \ref{sec:DC}. The EMS-based tertiary layer and its interaction with the secondary control layer is detailed in Section \ref{sec:EMS}. The in-depth functioning of secondary layer and related derivations are presented in Section \ref{sec:secondary}. \PNREV{Simulations provided in Section \ref{sec:simulations} substantiate theoretical results, and demonstrate the functioning and robustness of our control architecture on a modified 16-bus DC feeder \cite{Low}}, in the presence of inaccurate generation and load forecasts. Finally, conclusions are drawn in Section \ref{sec:conclusion}.
\begin{figure*}[h!]
	\centering
	\ctikzset{bipoles/length=0.85cm}
	\tikzstyle{every node}=[font=\scriptsize]
	\begin{circuitikz}[scale=0.67]
		\draw (1,1)  to [battery, o-o](1,4)
		to [short](1.5,4)
		to [short](1.5,4.5)
		to [short](3.5,4.5)
		to [short](3.5,0.5)
		to [short](1.5,0.5)
		to [short](1.5,4)
		to [short](1.5,1)
		to [short](1,1);
		\node at (2.5,3){ \textbf{DC-DC}};
		\node at (2.5,2.5){ \textbf{Converter}};
		\draw (3.5,4) to [short](4,4)
		to [short](4.5,4)
		to [R=$R_{ti}$] (6,4)
		to [L=$L_{ti}$] (7.5,4)
		to [short, -] (8.5,4)
		to [short](9,4); 
		\draw (9,1)
		to [short](4,1)
		to [short](3.5,1);
		\draw (8.5,4) to (11,4);
		\draw (9,1)
		to [short, -o] (15,1); 
		\draw (11,4) to [short](11.5,4);
		\draw (10,4) node[anchor=north, above]{$\textcolor{red}{V_i}$}  to [C, l=$C_{ti}$, o-] (10,1);
		\draw (10,4) to [short, i_=$\textcolor{red}{I_{i}}$](12,4) to [short] (15,4);
		\node at (10,4.6)[anchor=north, above]{$PC_i$} ;
		\draw (11,4) to (12.3,4); 
		\draw (15,4) to (18,4) to (21.5,4) 
		to [ I ] (21.5 ,1)
		to [short] (15,1)
		to [short, -] (21,1); 
		\draw (18.75,4) 
		to [C, l_=$C_{tj}$, o-] (18.75,1);
		\draw (11,4) to [short](17.5,4);
		\draw (21.5,4) to [short, i_=$I_{Lj}(V_j)$, -](21.5,2.9);
		\node at (18.75,4)[anchor=north, above]{$\PNREV{V_j}$} ;
		\node at (18.75,4.6)[anchor=north, above]{$PC_j$} ;
		\draw (16,4) -- (19,4); 
		\draw (18,4) to [short, i=$\PNREV{I_{j}}$](17.5,4);
		\draw[black, dashed] (19.75,.25) -- (22.25,.25) -- (22.25,5.2) -- (19.75,5.2)node[sloped, midway, above]{{ \textbf{Load $j$}}}  -- (19.75,5) -- (19.75,.25);
		\draw[black, dashed] (.5,.25) -- (9,.25) -- (9,5.2) -- (.5,5.2)node[sloped, midway, above]{{ \textbf{DGU $i$ }}}  -- (.5,.25);
		\node at (14.5,2.75) [cloud, draw, fill=blue!10!, line width=0.5mm, cloud puffs=10, cloud ignores aspect, cloud puff arc=110, minimum height=3.75cm] {\normalsize  DCmG Network};
	\end{circuitikz}
	\caption{Representative diagram of the DCmG network with DGUs and loads.}
	\label{fig:DGUnits}
\end{figure*}
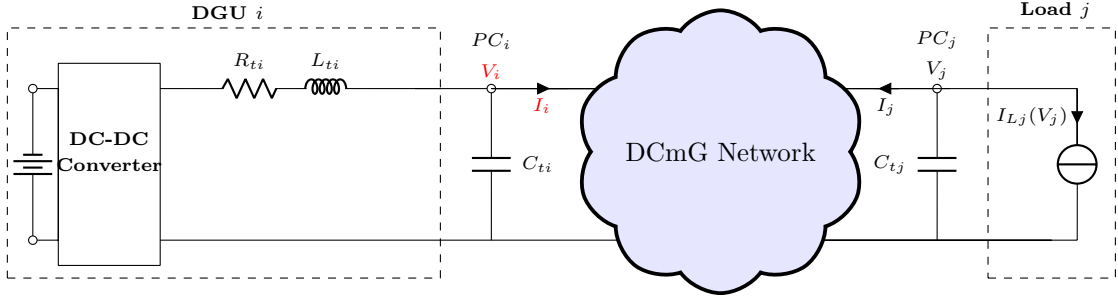

\subsection{Preliminaries and notation}
\textit{Sets, vectors, and functions:} We let $\mathbb{R}$ (resp. $\mathbb{R}_{>0}$) denote the set of real (resp. strictly positive real) numbers. Given $ x \in \mathbb{R}^{n}$, $[x] \in \mathbb{R}^{n \times n}$ is the associated diagonal matrix with $x$ on the diagonal. The inequality $x\leq y$ for vectors $x,y \in \mathbb{R}^{n}$ is component-wise, i.e. $x_i\leq y_i,~~\forall i\in 1,...,n$. For a finite set $\mathcal{V}$, let $|\mathcal{V}|$ denote its cardinality. Given a matrix $A \in \mathbb{R}^{n \times m}$, $(A)_i$ denotes the $i^{th}$ row. The notation $A \succ 0$ , $A \succeq 0$, $A>0$, and $A\geq 0$ represents a positive definite, positive semidefinite, positive, and nonnegative matrix, respectively.	Throughout, $\textbf{1}_n$ and $\textbf{0}_n$ are the $n$-dimensional vectors of unit and zero entries, and $\mbf{0}$ is a matrix of all zeros of appropriate dimensions.  

\PNREV{ \textit{Algebraic graph theory}: We denote by $\mathcal{G}(\mathcal{V},\mathcal{E},{W})$ an undirected graph, where $\mathcal{V}$ is the node set and $\mathcal{E}=(\mathcal{V}\times\mathcal{V})$ is the edge set. If a number $l \in \{1,...,|\mathcal{E}|\}$ and an arbitrary direction are assigned to each edge, the incidence matrix $B \in \mathbb{R}^{|\mathcal{V}|\times|\mathcal{E}|}$ has non-zero components: $B_{il} = 1$ if node $i$ is the source node of edge \textit{l}, and $B_{il} =-1$ if node $j$ is the sink node of edge $l$. The \textit{Kirchoff's Current Law} (KCL) can be represented as $x = B\xi$, where $x \in \mathbb{R}^{|\mathcal{V}|}$ and $\xi \in \mathbb{R}^{|\mathcal{E}|}$ respectively represent the nodal injections and edge flows. Assume that the edge $l \in \{1,...,|\mathcal{E}|\}$ is oriented from $i$ to $j$, then for any vector $V \in \mathbb{R}^{|\mathcal{V}|}$, $(B^TV)_l=V_i-V_j$. The Laplacian matrix $\mathcal{L}$ of graph $\GG$ is $\LL=BWB^T$, with $W$ being the diagonal matrix of edge weights.}
\section{DC microgrid structure and hierarchical control scheme}
\label{sec:DC}
In this section, we describe the DCmG structure and provide an outline of the hierarchical control structure used to ensure optimal, safe, and uninterrupted operation of the network.  

\textit{Structure of the DC microgrid:} \PNREV{A DCmG comprises multiple DGUs and loads, connected to each other via several power lines. Its electrical interconnections are modeled as an undirected connected graph $m\mathcal{G}=(\mathcal{V}, \mathcal{E})$, where $\mathcal{V}$ is the set of nodes and $\mathcal{E}$ the set of edges (power lines). The set $\VV$ is partitioned into two disjoint sets: $\mathcal{G}\subset \mathcal{V}$ is the set of DGUs and $\mathcal{L}\subset \mathcal{V}$ the set of loads. Figure \ref{fig:DGUnits} shows a representative diagram of the DCmG, highlighting the internal structure of DGUs and loads. It is worth noting that each DGU and load is interfaced with the rest of the DCmG (represented with a cloud in Figure \ref{fig:DGUnits}) via a Point of Coupling (PC), characterized by a voltage $V_i$ and output current $I_i$, with $i \in \mathcal{V}$.}

%

\textit{Distributed generation units (DGUs)}: \PNREV{A DGU consists of a DC voltage source,  a DC-DC converter, and a series $RLC$ filter (see Figure \ref{fig:DGUnits}; DGU filter capacitance is assumed to be lumped with capacitance $C_{ti}$). The DC voltage source can be of different types, such as dispatchable sources,  battery storage systems and PV panels. As the the type of DGU voltage source determines management strategies and system models to be adopted (see Section \ref{sec:EMS}), we define $\mathcal{G}_D$ as the set of DGUs interfaced with dispatchable sources, $\mathcal{G}_B$ as the set of DGUs interfaced with batteries, and $\mathcal{G}_P$ as the set of DGUs connected to PV panels, such that $\mathcal{G}_D\cup\mathcal{G}_B\cup\mathcal{G}_P =\mathcal{G}$.}

\textit{Load model:} The term $I_{Li}(V_i)$ in Figure \ref{fig:DGUnits} indicates the load's functional dependence on the PC voltage, and  takes different expressions based on the type of load. Prototypical load models that are of interest include the following:
\begin{enumerate}
	\item constant-current loads: $I_{LI,j} = \bar{I}_{L,j}$,
	\item constant-impedance loads: $I_{LZ,j}(V_j) = Y_{L,j}V_j$, where $Y_{L,j}= 1/R_{L,j}>0$ is the conductance of load $j$, and
	\item constant-power loads: 
	\begin{equation}
	\label{eq:constantPLoads}
	I_{LP,j}(V_j) = V_j^{-1}\bar{P}_{L,j},	
	\end{equation}		
	where $\bar{P}_{L,j} >0$ is the power demand of the $j^{th}$ load.
\end{enumerate}
To refer to the three load cases above, the abbreviations “I”, “Z”, and “P” are often used \cite{Kundur}. The analysis presented in this article will focus on the general case of a parallel combination of the three loads, thus on the case of “ZIP” loads, which are modeled as
\begin{equation}
\label{eq:loaddynamics}
I_{L,j}(V_j)=\bar{I}_{L,j} + Y_{L,j}V_j +V_j^{-1}\bar{P}_{L,j}.
\end{equation}
The net power absorbed by the $j^{th}$ load is given as
\begin{equation}
\hspace{-3mm}{P}_{L,j}(V_j)=\bar{I}_{L,j}V_j  + Y_{L,j}V^2_j +\bar{P}_{L,j}.
\label{eq:load_tot_power}
\end{equation}
\vspace{-5mm}
\subsection{Hierarchical control in DC microgrids}
In this work, we propose the hierarchical control architecture depicted in Figure \ref{fig:hierarchy}. The controller is split into three distinct layers \textit{viz.} primary, secondary, and tertiary. The secondary and tertiary layers together form the supervisory control layer of the DCmG network. 

All DGUs are equipped with local voltage regulators (not shown in Figure \ref{fig:DGUnits}) forming the \textit{primary control layer}. The main objective of these controllers is to ensure that the voltage at each DGU's \PNREV{PC}
tracks a reference voltage $V^*_{i}$ provided by the \textit{supervisory control layer}. 

\PNREV{\begin{assum}\textbf{(Stability under primary voltage control).}
In steady state, the primary controllers are assumed to achieve offset-free voltage tracking of constant voltage references $V^*_i, i \in \GG$, and to guarantee the stability of the entire DCmG network. 
\end{assum}
The above assumption implicitly states that, during transients and load variations, the primary-controlled DGUs never saturate, meaning that they have suffcient power at their disposal to feed the entirety of loads in the DCmG network.  For further details apropos of the design of stabilizing primary controllers, the reader is deferred to  \cite{Martinelli2018, Nahata, strehle2020scalable, cucuzzella2017decentralized} and the references therein.}

An EMS sits at the tertiary level, and utilizes the forecasts of PV generation $P_{PV}^{f}$, and load power and current absorption terms $\bar{P}^f_L, \;\bar{I}^f_L$. At each time step, it measures the nominal PV generation $P_{PV}^{o}$, the state of charge (SOC) of batteries ${S}_B$  
and the actual power and current absorption of ZIP loads  $\bar{P}_L,\bar{I}_L$. Solving an MPC optimization problem, the EMS generates optimal power references $\bar{P}_{G,i}, i \in \GG,$ for the DGUs. In addition, it produces decision variables $\delta_i \in \{0,1\}, i\in \GG$, which can either turn ON/OFF DGUs or change their operation mode. Since the primary layer operates only with voltage references, the secondary control layer translates the power references into appropriate voltage references $V^*$. The detailed structure and functioning of the secondary and tertiary control layers are discussed in Sections \ref{sec:secondary} and \ref{sec:EMS}, respectively. 

We highlight that different layers work at different time scales. In a typical scenario, the primary controllers  operates in a range varying from $10^{-6}$ to $10^{-3}$ s, the secondary layer ranges from $100$ to $300$ s, and the tertiary layer ranges from $5$ to $15$ mins \cite{Iovine2019}. At each high level sampling time, the controller  provides a reference to its corresponding lower layer.

\section{Tertiary control layer: the EMS}
\label{sec:EMS}
\PNREV{This section details the functioning of the MPC-based EMS occupying the topmost position in our proposed hierarchical structure.} The forecasts, parameters, and decision variables are described in Table \ref{tab:tab1}. As a convention, all the power values are defined to be positive if delivered from a DGU. Moreover, the upper and lower bounds of each variable are denoted with superscripts $max$ and $min$, respectively.
\begin{table}[t!]
	\caption{Optimization variables and parameters for the EMS}
	\label{tab:tab1}
	\centering
	\begin{tabular}{l l}
		\hline 
		Symbol  &Description\\ \hline \vspace{1mm}
		$P_{DH},P_{CH}$ & Charging and discharging power of the battery [kW]\\ \vspace{1mm}
		$P_{B}$   &   Power output of battery DGUs [kW] \\ \vspace{1mm}
		$P_{D}$   & Power output of dispatchable DGUs [kW]\\ \vspace{1mm}
		$P_{PV}$ & Power output of PV DGUs [kW]\\ \vspace{1mm}
		$P_{PV}^o$ & Nominal power production of PV DGUs [kW]\\ \vspace{1mm}
		$P_{PV}^{f}$   & Power production forecast of PV DGUs [kW]\\ \vspace{1mm}
		${P}^o_{L}$      &  Nominal total power absorption for {ZIP loads} [kW]\\ \vspace{1mm}
		$\bar{I}^f_L$ & Current absorption forecast for {I load} [kVar]\\ \vspace{1mm}
		$\bar{P}^f_{L}$ & Power absorption forecast for {P load} [kW]\\ \vspace{1mm}
		\(S_{B}\) & State of charge (SOC) of battery  \\ \vspace{1mm}
		\(S_{B}^o\) &Nominal SOC of battery \\ \vspace{1mm}
		$\eta_{CH},\eta_{DH}$ &Charging and discharging efficiency of battery  \\ \vspace{1mm}
		$ C_{B} $ & MG battery capacity [kWh] \\ \vspace{1mm}
		\( V^o \)         & Nominal network voltage [V]\\ \vspace{1mm}
		$ \delta_{B} $ &  Operation mode of battery DGU [boolean]\\ \vspace{1mm}
		$ \delta_{D} $ & Operation mode of dispatachable DGU [boolean] \\ \vspace{1mm}
		$\delta_{PV}$    & Operation mode of PV DGU [boolean]\\ \vspace{1mm}
		\(V  \) & Nodal voltage magnitude  [V]\\ \vspace{1mm}
		$I$ & Nodal current magnitude [A] \\
		\hline
	\end{tabular}
\vspace*{-3mm}
\end{table}

\subsection{MPC-based EMS for islanded DCmGs}\label{subsec:mg_model}
The MPC--based EMS is responsible for energy management and coordination of resources in the islanded DCmG. At the core of this controller is a receding horizon optimization problem which enables load satisfiability, optimal scheduling of dispatchable and storage DGUs, and maximum possible utilization of PV DGUs. 

\PNREV{ At a generic time instant $k$, the EMS solves a mixed integer optimization problem over a finite prediction horizon $ [ k, \hdots ,k+N ] $, with $N$ indicating the number of prediction steps. As a consequence,  an optimal plan (input) on power dispatch, storage schedule, and operational modes of DGUs is formulated for the entire prediction horizon. Nevertheless, only the first sample of the input sequence is implemented, following which the horizon is shifted. At the next sampling time, using updated information on forecasts and mG initial condition, the EMS solves a new optimization problem. }

\PNREV{In the ensuing discussion, we lay out the EMS optimization problem in detail. Unless stated otherwise, the index $i$ spans the entire prediction horizon except for the terminal time step $N$, i.e. $i\in[0,\hdots,N-1]$.} 

\begin{enumerate}[leftmargin=*]
	
	\item \textit{DGUs}: \PNREV{For the purpose of the EMS optimization problem, the characterization of a DGU hinges on the type of its voltage source.}
	\begin{itemize}[leftmargin=*]
		\item[a)] Storage DGUs: 
		\PNREV{A battery serves as the voltage source for these DGUs.
		Accounting for both the charging and discharging efficiencies, the SOC dynamics of a battery $b\in\GG_B$  are given as}
		\begin{equation}
		\begin{split}
		&S_{B,b}(k+1+i)  = S_{B,b}(k+i)\;-  \\
		&\frac{\tau}{C_{B,b} } \Bigg( \frac{1}{\eta_{DH,b}}P_{DH,b}(k+i) +  {\eta_{CH,b}} P_{CH,b}(k+i)\Bigg),
		\label{SOC_dyn}
		\end{split}
		\end{equation}
		with battery power output
		\begin{equation}
		\label{eq:batterypoweroutput}
		P_{B,b}(k+i)=P_{DH,b}(k+i) \text{\PNREV{\,+\,}} P_{CH,b}(k+i).
		\end{equation}
		Since battery DGUs can operate either in charging or in discharging mode, the following constraints are stated
		\begin{align}
		0 \, \leq  &P_{DH,b}(k+i)\leq{P}^{{\,max}}_{B,b}(k+i)\, \delta_{B,b}(k+i), \label{discharge_limits}  \\
		0 \, \leq &P_{CH,b}(k+i) \leq -{P}^{{\,min}}_{B,b}(k+i)\,  (1-\delta_{B,b}(k+i)), \label{charging_limits}
		\end{align}		
		where $\delta_{B,i}=1$ indicates discharging mode while $\delta_{B,i}=0$ represents the charging mode.  
		In order to ensure longevity of batteries, we constrain the SOC between minimum and maximum bounds as
		\begin{align}
		\begin{split}
		{S}^{{\,min}}_{\,B,b} \, \leq \; {S}_{\,B,b}&(k+i) \, \leq \,  {S}^{{\,max}}_{\,B,b}.   \\
		\end{split}
		\label{SOC_constr}
		\end{align}
\PNREV{	To avoid complete charging or discharging of batteries---not ideal for guaranteeing uninterrupted power supply to loads in face of a contingency, we constraint the terminal SOC as
		\begin{equation}
		\label{eq:SOCterminalconstraint}
		{S}_{\,B,b}(k+N) =\;{S}^{o}_{\,B,b}+\Delta{S}_{\,B,b}\, ,
		\end{equation}
		where $ {S}^{o}_{\,B,b}$ is the nominal SOC of battery $b\in\GG_B$, while $\Delta{S}_{\,B,b} $ is a slack variable introduced to ensure feasibility of the EMS optimization problem.}
		\item[b)] Dispatchable DGUs:  \PNREV{Equipped with a dispatchable voltage source---commonly nonrenewable, e.g. a fuel cell or a DC electric generator, these DGUs can be switched ON/OFF, based on need. Their operation mode is governed by the variable $\delta_{D,d},d\in\DD_\DD$, with values $1$ and $0$ indicating ON and OFF states, respectively. }The power produced lies within a range defined by lower and upper bounds
		\begin{equation}
		\begin{split}
		&\delta_{D,d}(k+i)\,{P}^{{\,min}}_{D,j}	\;\leq \;P_{D,d}(k+i)\, \\ 
		&\;P_{D,d}(k+i)\, \leq \,	{P}^{{\,max}}_{D,d} \,\delta_{D,d}(k+i)	,\;\; ~d\in\GG_D \label{dispatchable_limits}  
		\end{split}.
		\end{equation}	

		\item[c)] PV DGUs: \PNREV{The $p^{th}$ DGU, $p \in \mathcal{G}_P$, has two distinct modes of operation:  power curtailment mode and  MPPT. PV DGUs inject maximum available power into the grid while operating in MPPT mode. They otherwise curtail power---unavoidable during periods of peak PV generation---to preserve internal power balance of the DCmG. Since, at a given time instant, the EMS utilizes both actual nominal PV production and future PV forecast, the PV power output is expressed as
		\begin{align}
		P_{PV,p}(k)  &={P}^o_{PV,p}(k) + \Delta {P}_{PV,p}(k), \label{eq:Pvcurrent} \\
		P_{PV,p}(k+i) &= {P}^f_{PV,p}(k+i)-\Delta {P}_{PV,p}(k+i),  \label{eq:Pvforecast} 
		\end{align}		
where $\Delta {P}_{PV,p}$ expresses the amount of curtailed power. Additionally, $\Delta {P}_{PV,p}$ fulfills 
		\begin{align}
		\Delta {P}_{PV,p}(k) &\geq (1-\delta_{PV,p}(k) )\, \epsilon,\label{PV_limits_current}	\\ 
		\Delta {P}_{PV,p}(k) &\leq (1-\delta_{PV,p}(k) )\,{P}^o_{PV,p}(k),  \\[2mm] 
		\Delta {P}_{PV,p}(k+i) &\geq (1-\delta_{PV,p}(k+i) )\epsilon, 	\\ 
		\Delta {P}_{PV,p}(k+i) &\leq (1-\delta_{PV,p}(k+i) )\,{P}^f_{PV,p}(k+i),
		\label{PV_limits_forecast} 
		\end{align}
 where $\epsilon>0$ is a sufficiently small number and $\delta_{PV,p}$ is a decision variable. The rationale behind constraints \eqref{PV_limits_current}-\eqref{PV_limits_forecast} is not only to limit power curtailment bound by the nominal PV production, but also to allow just one of the operation modes at a time.  Clearly, if $\delta_{PV,p}=1$, $\Delta {P}_{PV,p}$ is forced to zero  meaning that the MPPT mode is activated, whereas if $\delta_{PV,p}=0$, the curtailed power must be strictly greater than zero and lower than the nominal PV power production. For more details on logic and mixed-integer constraints, the reader is deferred to \cite{BemporadMPC}. }
	\end{itemize} 
	\item \textit{Loads:} \PNREV{At $t=k$, the nominal power absorption of the $l^{th}$ ZIP load, $l\in\LL$ for the first time step is computed at nominal voltage by means of the current state of the system
	\begin{equation}
	\label{eq:loadconsumption}
	P^o_{L,l}(k)=\bar{I}_{L,l}(k)V^o + Y_{L,l}{V^o}^2 +\bar{P}_{L,l}(k),~ l \in\LL.
	\end{equation}
Load forecasts are used for future time steps \linebreak $i \in [1,\hdots, N-1]$. Therefore,}
	\begin{equation}
	\label{eq:loadfconsumption}
	P^o_{L,l}(k+i)=\bar{I}^f_{L,l}(k+i)V^o + Y_{L,l}{V^o}^2 +\bar{P}^f_{L,l}(k+i).
\end{equation}  
	It is worth noticing that $P^o_{L,l}$ is just an estimate since net power absorption of ZIP loads depends on the actual DCmG voltages, see \eqref{eq:load_tot_power}.
	\item \textit {Power balance:} In an islanded DCmG, the internal power balance must be maintained. Hence, the following constraint is introduced
	\begin{equation}
	\begin{split}
	\sum_{b\in\DD_\BB}&P_{B,b}(k+i) + \sum_{d\in\DD_\DD}P_{D,d}(k+i) \\ + &\sum_{p\in \DD_\PP}P_{PV,p}(k+i) +\sum_{l\in \LL}P_{L,l}^{o}(k+i)=0, \label{mg_output_power}
	\end{split}.
	\end{equation}
	 We highlight that the converter and network losses are neglected at the EMS level.
	
	\item \textit{Cost function:} \PNREV{Our aim is to minimize the cost of satisfying the electrical loads; hence the cost function is}
	\begin{align}
	\begin{split}
	&J(k) = \\	+&{\sum_{b\in\DD_\BB}(\Delta{S}_{B,b})^2}w_{S,b}+ \sum_{i=0}^{N-1}{\sum_{b\in\DD_\BB}(P_{B,b}(k+i))^2}w_{B,b}  \\ 
	+ & \sum_{i=0}^{N-1}{\sum_{d\in\DD_\DD}(P_{D,d}(k+i))^2}w_{D,d}  \\
	+&\sum_{i=0}^{N-1}{\sum_{p\in\DD_\PP}( \Delta {P}_{PV,p}(k+i))^2}\,w_{PV,p}    \\  	+&\underbrace{\sum_{i=0}^{N-1}{\sum_{p\in\DD_\PP}(\delta_{PV,p}(k+i)-\delta_{PV,p}(k+i-1))^2}w\delta_{PV,p}}_{\alpha}\\
	+&\underbrace{\sum_{i=0}^{N-1}{\sum_{b\in\DD_\BB}(\delta_{B,b}(k+i)-\delta_{B,b}(k+i-1))^2}w\delta_{B,b}}_{\beta}\\ +&\underbrace{\sum_{i=0}^{N-1}{\sum_{d\in\DD_\DD}(\delta_{D,d}(k+i)-\delta_{D,d}(k+i-1))^2}w\delta_{D,d}}_{\gamma} \quad,\\
	\end{split}
	\label{muG_costfunction}
	\end{align}
\PNREV{where $w_{S}, w_{PV}, \dots $ are positive weights. In particular, $w_{S}$ weighs the slack variable $\Delta{S}_{B,b}$, whereas $w_{B},\, w_D$ weigh the power outputs of battery and dispatchable DGUs, respectively. We intend to keep batteries close to their nominal SOCs, and to use power curtailment as the last resort. Thus, the weights $w_{S,B}$ and $w_{PV}$ are set to much higher values with respect to others, permitting $\Delta{S}_{\,B,b}$ and $\Delta {P}_{PV,p}$ to be nonzero only when necessary for preserving feasibility. The terms $\alpha$, $\beta$ and $\gamma$ are included in the cost to avoid frequent changes in modes of operation of different DGUs. \\}
\end{enumerate}

At every EMS time instant, the following optimization is solved to obtain optimal power set points $\bar{P}_{B,i}, \bar{P}_{D,j}, \bar{P}_{PV,p}$ and decision variables $\delta_{B,i}, \delta_{D,j}, \delta_{PV,p}$.       
\begin{subequations}
	\begin{align}
	\hspace{-15mm}J_{\scriptscriptstyle {EMS}} (k) =&  \min J(k)
	\label{eq:EMSsPF_cf}\\
	&\hspace{-15mm} \text{subject to} \nonumber\\
	&\quad \eqref{SOC_dyn}-\eqref{mg_output_power}.
	\end{align}\label{eq:EMSsPF}
\end{subequations}
\vspace*{-10mm}
\subsection{Interaction between tertiary and secondary layers}
The EMS produces power references as well as decision variables, both of which are passed down to the secondary control layer. \PNREV{Given that  the decision variable $\delta_{D,j}$ decides whether a dispatchable DGU gets connected to, or disconnected from, the DCmG network, the EMS essentially determines the topology of the DCmG network. Moreover, the PV DGUs can either inject maximum power or undergo power curtailment, depending upon the value of $\delta_{PV,p}$. While injecting maximum power, the PV DGU governed by standard MPPT algorithms automatically alters its output voltage to inject maximum power. Thus, in this mode, the PV DGU mimics a P load injecting power. While experiencing a power curtailment, the PV DGU operates as a voltage-controlled DGU and injects the requested power.}

\PNREV{As stated earlier, the EMS power references cannot be perceived by the primary voltage controllers.  To carry out a power-to-voltage translation, the intermediary secondary layer solves an optimization problem relying on power-flow equations (see Section \ref{sec:secondary}).  Since these equations are dependent on DCmG topology and parameters, the secondary controller, at every EMS time instant, is required to exploit the decision variables to account for the operation mode of PV DGUs and the changes in DCmG topology caused caused by dispatchable DGUs. }
\begin{remark}\textbf{(Connectivity of the DCmG network).} \PNREV{Turning ON/OFF dispatchable DGUs is assumed not to impact the connectivity of the rest of the DCmG network.} In other words, addition or removal of a dispatchable  DGUs must not split the remainder of the network into two or more disjoint islanded mGs. In case critical DGUs affecting the connectivity of graph are present in the network, one can restrict their operation modes by adding additional constraints to the EMS optimization problem (see Section \ref{sec:simulations} for an example).
\end{remark}
\section{Secondary control based on power-flow equations}
\label{sec:secondary}
\PNREV{The secondary control is designed to facilitate DGUs' generating EMS power references, denoted as $\bar{P}_G$ from now on. We recall that the decision variables communicated at an EMS sampling instant define the topology of the network over the course of the successive EMS sampling period.}
\begin{remark}
	The secondary layer, operating on a faster time scale in comparison to the EMS, utilizes a fixed DCmG topology over an EMS sampling period to perform power-voltage translation. The topology is updated when a new set of decision variables is received. 
\end{remark}
\PNREV{To generate proper primary voltage references out of EMS voltage references, the secondary control layer needs to link powers and voltages. Therefore, we start by deducing the relation between powers and voltages, defined by the power-flow equations dependent on DCmG parameters and topology.}

We let the undirected connected graph $m\mathcal{\tilde{G}}=(\mathcal{\tilde{V}}, \mathcal{\tilde{E}})$ define the topology of the DCmG for a specified EMS sampling period. The set $\tilde{\VV}$ is partitioned into two sets: $\tilde{\GG}=\{1,\dots,n\}$ is the set of DGUs and $\tilde{\LL}=\{n+1,\dots,n+m\}$ the set of loads. The set $ \tilde{\GG}=\tilde{\GG}_D\cup\tilde{\GG}_B\cup\tilde{\GG}^G_P$, where $\tilde{\GG}_D$ is the set of connected dispatchable DGUs, $\tilde{\GG}_B$ the set of batteries, and $\tilde{\GG}_P^D$ the set of voltage-controlled PV DGUs. In steady state, the inductances and capacitances can be neglected. Hence, the current-voltage relation is given by the identity $ I = B{\Gamma}B^TV=YV$, where, $I$ is the vector of \PNREV{PC} output currents, $V$ the vector containing \PNREV{PC} voltages (see Figure \ref{fig:DGUnits}), $\Gamma$  the diagonal matrix of line conductances, and  $Y\in \mathbb{R}^{(n+m) \times (n+m)}$ the network admittance matrix \cite{dorfler2018electrical}. \PNREV{On partitioning the nodes into DGUs and loads, one obtains the above relation as}
\begin{equation}
\label{eq:Laplacian}
\begin{split}
\begin{bmatrix}
I_G\\
I_L
\end{bmatrix}
&=\begin{bmatrix}
B_GR^{-1}B_G^T &B_GR^{-1}B_G^T\\
B_LR^{-1}B_G^T &B_LR^{-1}B_G^T
\end{bmatrix}\begin{bmatrix}
V_G\\
V_L
\end{bmatrix}\\
&:=
\begin{bmatrix}
Y_{GG} &Y_{GL}\\
Y_{LG} &Y_{LL}
\end{bmatrix}\begin{bmatrix}
V_G\\
V_L
\end{bmatrix}\\
\end{split},
\end{equation}
where $V_G=[V_1,\dots , V_n]^T$, $V_L=[V_{n+1},\dots , V_{n+m}]^T$, \mbox{$I_G=[I_1,\dots , I_n]^T$}, and \mbox{$I_L=[I_{n+1},\dots , I_{n+m}]^T$}. Moreover, \PNREV{\mbox{${Y}_{GG} \in \mathbb{R}^{n \times n}$}, \mbox{${Y}_{GL} \in \mathbb{R}^{n \times m}$}, \mbox{${Y}_{LG} \in \mathbb{R}^{m \times n}$} and  \mbox{$Y_{LL} \in \mathbb{R}^{m \times m}$}}. The subscripts $G$ and $L$ indicate the DGUs and loads, respectively. 
Throughout this work, the following assumption is made.

\smallskip 
\begin{assum}
	The 
	\PNREV{PC} voltage $V_i$ is strictly positive for all $i\in\VV$.	 
	\label{ass:Vpos}
\end{assum}
We remark that Assumption \ref{ass:Vpos} is not a limitation, and rather reflects a common constraint in microgrid operation.  Notice that, in Figure \ref{fig:DGUnits}, one end of the load is connected to \PNREV{the PC}
 and the other to the ground assumed be at zero potential by convention. Since the electric current flows from higher to lower potential,  
\PNREV{negative PC voltages} will reverse the role of loads and make them power generators. In order to ensure power balance in the network, the DGUs will have to absorb this surplus power. This, in effect, defeats the fundamental goal of the mG, i.e. the satisfiability of the loads by virtue of the power generated by the DGUs. Furthermore, if $V_i \in \mathbb{R}^N$, then a zero-crossing for the voltages may take place. At zero voltage, the power consumed by the ZIP loads tends to infinity. 

Based on the current directions depicted in Figure \ref{fig:DGUnits}, $I_{L,j}(V_j)=-I_j,~j\in\LL$. Using \eqref{eq:loaddynamics}, one can simplify \eqref{eq:Laplacian} as 
\begin{subequations}
	\begin{align}
	&I_G=Y_{GG}V_{G} \,+\,  Y_{GL} \,V_L  \label{eq:It_equilibrium}
	\\
	&0=Y_{LG}V_{G}  \,+\,  Y_{LL}\,V_L  +Y_L V_{L}+\bar{I}_L+ \, [V_L]^{-1} \, \bar{P}_L,
	\end{align}
	\label{eq:equilibriumstate}
\end{subequations}
where $Y_L \in \mathbb{R}^{m\times m}$ is the diagonal matrix of load admittances. The vectors $\bar{I}_L$ and $\bar{P}_L$ collect consumptions of I and P loads, respectively.
Note that the power $P_{G,i}, i \in \tilde{\GG}$ produced by an individual DGU is the sum of power injected into the network and the filter losses. Equivalently, 
\begin{equation}\label{eq:pg}
P_G=[V_G]I_G+[I_G]R_{G}I_G
\end{equation}
where $R_G \in \mathbb{R}^{n\times n}$ is a diagonal matrix collecting filter resistances and $I_G$ is the vector of DGU filter currents. On pre-multiplying \eqref{eq:It_equilibrium} with $[V_G]$, and by using \eqref{eq:pg}, one can rewrite \eqref{eq:equilibriumstate} as  
\begin{equation}
\label{eq:f_G}
\begin{split}
f_G (V_G,V_L,P_G) &=[V_G] {Y}_{GG} \,V_G \,+\,  [V_G] {Y}_{GL} \,V_L \,   \\
& \,+ \, [I_G] \,R_G \, I_G \,-\, P_G =0,
\end{split}
\end{equation}
\begin{equation}
\label{eq:f_L}
\begin{split}
f_L (V_G,V_L) &= Y_{LG} \,V_G \, +Y_{LL} \,V_L +Y_LV_L \, \\&+Y_L V_L+\bar{I}_L+ \, [V_L]^{-1} \, \bar{P}_L=0. 
\end{split}
\end{equation}
Equations \eqref{eq:f_G} and \eqref{eq:f_L} fundamentally depict the power and current flow at DGU and load nodes, respectively. These equations depend on the topology-dependent $Y$ matrix, and are updated once a new set of decision variable is received. 

In order to translate the power references into suitable voltage references, the secondary layer solves the optimization problem \eqref{eq:CPF}, whose objective is to minimize the difference between the reference power $\bar{P}_G$ and the DGU input power $P_G$ under the equilibrium relations \eqref{eq:f_G} and \eqref{eq:f_L}, \PNREV{as well as DGU capability limits and DCmG voltage constraints. Before presenting \eqref{eq:CPF}, we address the solvability of the nonlinear, non-convex equations \eqref{eq:f_G} and \eqref{eq:f_L}. On this account, we first consider the following simplified version of the optimization problem \eqref{eq:CPF}, where nodal voltages and generator power are not bounded.}
\\ 
\\
\textbf{Secondary Power Flow  (SPF)}: 	        
\begin{subequations}
	\begin{align}
	\hspace{-15mm}J_{\scriptscriptstyle {SPF}} (\bar{P}_G,\bar{P}_L,\bar{I}_L) =&  \min_{\substack{ \,V_G,\, V_L, P_G}} \;||P_G -\bar{P}_G||_2 
	\label{eq:sPF_cf}\\
	&\hspace{-15mm} \text{subject to} \nonumber\\
	&\quad f_G (V_G,V_L,P_G)  = 0\label{eq:sPF_G}
	\\
	&\quad f_L (V_G,V_L)  = 0\label{eq:sPF_L}
	\end{align}\label{eq:sPF}
\end{subequations}
As noticeable from Figure \ref{fig:hierarchy}, the SPF layer requires the updated load consumption $(\bar{P}_L,\bar{I}_L)$ and the power references $\bar{P}_G$ in order to solve \eqref{eq:sPF}.
We define $\XX$ to be the set of all $(V_G,V_L,P_G)$ that satisfy \eqref{eq:sPF_G}-\eqref{eq:sPF_L} simultaneously. \PNREV{ Next we show that the set $\XX$ is nonempty, i.e. \textbf{SPF} is always feasible.}

\smallskip
\begin{proposition}\label{prop:existence}
	\textbf{(Feasibility of SPF).} The feasible set $\XX$ is non-empty . In particular, for all $\bar{P}_L \in \mathbb{R}^m$ and $\bar{I}_L \in \mathbb{R}^m$, the following statements hold:
	\begin{enumerate}
		\item The equation \eqref{eq:sPF_L} is always solvable. 
		\item The solvability of \eqref{eq:sPF_L} implies that \eqref{eq:sPF_G} is solvable.
	\end{enumerate}
\end{proposition}
\smallskip
\begin{proof}
	Under Assumption \ref{ass:Vpos}, \PNREV{multiplying \eqref{eq:sPF_L} with $[V_L]$ gives
		\begin{equation}
		[V_L]\,{Y}_{LG} \,V_G + 	[V_L]\,\tilde{Y}_{LL} \,V_L  +[V_L]\,\bar{I}_L+ \bar{P}_L =0\,,
		\label{eq:f_L_q}
		\end{equation}}	
where  $\tilde{Y}_{LL}= {Y}_{LL}+ {Y}_{L}$. As shown in \cite{simpson2016voltage}, using Banach fixed-point theorem,  one can prove that, for a  fixed ${V}_G$, a corresponding $V_L$ solving  \eqref{eq:sPF_L} exists if 
	\begin{equation}
	\label{eq:SPcond}
	\Delta = ||P^{-1}_{crit} \, \bar{P}_L||_{\infty} \; < \;1
	\end{equation}
	where 
	\begin{equation}
	P_{crit}= \frac{1}{4} \,[\tilde{V}]\, \tilde{Y}_{LL}\, [\tilde{V}]
	\label{eq:Pcrit}
	\end{equation}
	and
	\begin{equation}
	\tilde{V} = - \tilde{Y}^{-1}_{LL} \,{Y}_{LG}V_G - \tilde{Y}^{-1}_{LL} \bar{I}_L,
	\label{eq:Vtilde}
	\end{equation}
	\PNREV{where \mbox{$\tilde{Y}_{LL} \in \mathbb{R}^{m \times m}$} and \mbox{$Y_{LG} \in \mathbb{R}^{m \times n}$}}.
	Different from fixed ${V}_G$ in \cite{simpson2016voltage}, ${V}_G$ here is a free variable. Therefore, for \eqref{eq:sPF_L} to be solvable,  it is enough to show that a ${V}_G$ can be always found such that \eqref{eq:SPcond} is satisfied for any $\bar{I}_L$ and $\bar{P}_L$. \\
	Consider $V_G^{\alpha} = \alpha \, \mathbf{1}_{n}$, with $\alpha \in \mathbb{R}_{>0}$. Therefore,
	\begin{equation}
	\tilde{V}^{\alpha}= - \tilde{Y}^{-1}_{LL} \,{Y}_{LG}V_G^{\alpha} - \tilde{Y}^{-1}_{LL} \bar{I}_L =  \alpha (-\tilde{Y}^{-1}_{LL} \,{Y}_{LG} \mathbf{1}_{n}) - \tilde{Y}^{-1}_{LL} \bar{I}_L. 
	\label{eq:v_alpha}
	\end{equation}
	Given Lemma \ref{lemma:yinv_ylg}, $(-\tilde{Y}^{-1}_{LL} \,{Y}_{LG} \mathbf{1}_{n})$ is a positive vector. Hence, there exists an $\bar{\alpha} \in \mathbb{R}_{>0}$ such that $\tilde{V}^{\alpha}>0$ \, $ \forall \alpha>\bar{\alpha}$.
	\\
	\PNREV{At this stage, substituting $\tilde{V}^{\alpha}$ in \eqref{eq:Pcrit} leads to
		\begin{equation}
		P_{crit}^{\alpha}= \frac{1}{4} \,[	\tilde{V}^{\alpha}]\, \tilde{Y}_{LL}\, [	\tilde{V}^{\alpha}]\,.
		\label{eq:Pcrit2}
		\end{equation}
		Equivalently, 
		\begin{equation}
		(P_{crit}^{\alpha})^{-1}= 4 \,[	\tilde{V}^{\alpha}]^{-1}\, \tilde{Y}_{LL}^{-1}\, [	\tilde{V}^{\alpha}]^{-1}\,,
		\label{eq:Pcrit3}
		\end{equation}
		Any element $(i,j)$ of the matrix $(P^{\,\alpha}_{crit})^{-1}$, with   $i,j \in \mathcal{L} $ can be expressed as 
		\begin{equation}
		(P^{\,\alpha}_{crit})^{-1}_{ij} \, = \, 4 \,{(\tilde{Y}_{LL})^{-1}_{i,j}}/({\tilde{V}^{\alpha}_i \; \tilde{V}^{\alpha}_j}).
		\label{eq:Pcrit4}
		\end{equation}
		Considering \eqref{eq:v_alpha} and \eqref{eq:Pcrit4}, we conclude that $(P^{\,\alpha}_{crit})^{-1}_{ij}$ is inversely proportional to the parameter $\alpha$, for $\alpha > \bar{\alpha}$.\;}
	
	As a result, it is always possible to increase $\alpha$ such that \eqref{eq:SPcond} is verified for any $\bar{P}_L$ and $\bar{I}_L$.  Consequently, a voltage solution $(V_G^*,V_L^*)$ of \eqref{eq:sPF_L} always exists, proving statement 1.\\
	As to statement 2, it is clear that \eqref{eq:sPF_G} is linear with respect to $P_G$. This implies that, for any solution $(V_G^*,V_L^*)$ of \eqref{eq:sPF_L}, a corresponding $P_G^*$ solving  \eqref{eq:sPF_G} always exists.
\end{proof}
In a real DCmG, the power output $P_G$ is constrained by physical capability limits of the DGUs. Moreover, the components of the DCmG are designed to operate around the nominal voltage. Hence, both nodal voltages and DGU powers must respect certain bounds, which are not incorporated in the aforementioned \textbf{SPF}. Consequently, we now introduce the following constrained optimization problem with additional operational constraints.
%
%
\\ 
\\
\textbf{Secondary Constrained Power Flow  (SCPF)}: 	        
\begin{subequations}
	\begin{align}
	\hspace{-15mm}J_{\scriptscriptstyle {SCPF}} (\bar{P}_G,\bar{P}_L,\bar{I}_L) \;=\;& \min_{\substack{ \,V_G,\, V_L, P_G}} \;||P_G -\bar{P}_G||_2 
	\label{eq:PF_cf}\\
	&\hspace{-15mm} \text{subject to} \nonumber\\
	&\quad f_G (V_G,V_L,P_G)  = 0\label{eq:CPF_G}
	\\
	&\quad f_L (V_G,V_L)  = 0\label{eq:CPF_L}\\
	&\quad V_G^{min} \leq V_G  \leq V_G^{max}\label{eq:CPF_VGconstr} \\
	&\quad V_L^{min} \leq V_L  \leq V_L^{max}\label{eq:CPF_VLconstr} \\
	&\quad P_G^{min} \leq  P_G  \leq  P_G^{max} \label{eq:CPF_Pconstr}
	\end{align}\label{eq:CPF}
\end{subequations}
The feasibility of \textbf{SPF}, corresponding to solving \eqref{eq:PF_cf}-\eqref{eq:CPF_L}, is already ensured by Proposition \ref{prop:existence}. Considering the voltages and power bounds \eqref{eq:CPF_VGconstr}-\eqref{eq:CPF_Pconstr}, the overall feasibility of \textbf{SCPF} is not a priori guaranteed. Nevertheless, if the DCmG is properly designed, a feasible solution of \textbf{SCPF} should always exist. 
In fact, the infeasibility of the \textbf{SCPF} just implies the absence of sufficient power generation to satisfy the load demand and losses in the allowed voltage range.
\smallskip
 If \textbf{SCPF} achieves the optimal cost $J_{\scriptscriptstyle {SCPF}}^* =0$, \PNREV{it implies that a voltage solution corresponding to the power references $\bar{P}_G$ exists.} This condition can not be achieved for any value of $(\bar{P}_L,\bar{I}_L,\bar{P}_G)$. The following proposition, inspired by \cite{sanchez2014conditions}, presents a necessary condition that must hold when $J_{\scriptscriptstyle {SCPF}}^* =0$. The proof nonetheless is different as here DGU filter losses are also taken into account.
\smallskip
\begin{proposition}
	If \textbf{SCPF} achieves the optimal cost $J^*_{\scriptscriptstyle SCPF} =0$, then 
	\begin{align}
	&\sum_{\forall i \in \DD} \bar{P}_G \geq \sum_{\forall i \in \LL} \bar{P}_L - \frac{1}{4}\, \bar{I}_L^{\;T}\,  \tilde{Y}^{-1}_{GG} \,\bar{I}_L ,
	\label{eq:spf_neccond}
	\end{align}
	where  $\tilde{Y}_{GG}={Y}_{GG} \,-\, {Y}_{GL}^T ({Y}_{LL}+ {Y}_{L}) {Y}_{GL}$.
\end{proposition}
\smallskip
\begin{proof}
\PNREV{The proof is provided in Appendix.}
\end{proof}
\PNREV{
\begin{remark}
The necessary condition \eqref{eq:spf_neccond} depends only on
the network parameters and load consumption, and can be
incorporated in the EMS optimization problem as a constraint.
\end{remark}}
\PNREV{
\begin{remark}
\label{rem:trackingsamplingtimes}
Provided that the optimal cost of SPF/SCPF is zero, the EMS reference power $\bar{P}_G$ is that effectively produced by DGUs at a secondary sampling instant. Indeed, the loads may change between two subsequent sampling times. Since DGU voltages remain fixed over the course of a sampling interval, the DGUs are obliged to change their power generation in order to maintain DCmG power balance, leading to a deviation from the EMS powers. If the cost at the next sampling instant is zero, $\bar{P}_G$ is reinstated in the DCmG (consult Section \ref{sec:simulations} for an example).
\end{remark}}
Next we study the properties of an optimal solution \linebreak $\textbf{x}^*=(V^*_G,V^*_L,P^*_G)$ of \textbf{SCPF}, assuming it exists.
As stated previously, the secondary control layer acts as an interface between the EMS (tertiary layer) and the local voltage regulators (primary layer). The voltage $V^*_G$ obtained from the \textbf{SCPF} is transmitted as a reference to the primary voltage controllers of the DGUs. We highlight that just the component  $V^*_G$ of $\textbf{x}^*$ can be imposed directly since the load nodes are not equipped with voltage controllers and the generators are not controlled to work on power references. Therefore, it is important to guarantee that, for a given voltage reference $V^*_G$ at DGU nodes, $P^*_G$ is the power effectively produced and $V^*_L$ appears at the load nodes. This implies that for a fixed $V^*_G$, the unique solution satisfying the power flow equation \eqref{eq:f_G}-\eqref{eq:f_L} must be $V_L=V^*_L, \,P_G=P^*_G$. We show this by means of the following theorem.
\smallskip
\begin{theorem} \textbf{(Uniqueness of a voltage solution).}
	Consider the solution $\textbf{x}^*=(V^*_G,V^*_L,P^*_G)$ from the \textbf{SCPF} optimization problem. For a fixed $V^*_G$, the pair  $(V^*_L,P^*_G)$ is the unique solution of \eqref{eq:f_G}-\eqref{eq:f_L} in the set \linebreak $\YY =\{ (V_L, P_G) :  V_L  >  V_L^{min}, \; P_G \in \mathbb{R}^n\}$ if
	\begin{equation}
	\label{eq:PLcond}
	\bar{P}_{L,i} < (V_{i}^{min})^2 \, {Y}_{L,i}, \quad \forall i \in \tilde{\LL}.
	\end{equation}
\end{theorem}

\smallskip
\begin{proof}
	For a fixed $V_G^*$, the power-flow equations \eqref{eq:f_G}-\eqref{eq:f_L} can be rewritten as
	\begin{align}
	\begin{split}
	\tilde{f}_G (V_L,P_G)&= \; f_G (V_G,V_L,P_G)\bigg|_{V_G=V_G^*} \;=\;[V_G^*] {Y}_{GG}V^*_G \\&+ \; [V_L] {Y}_{LG} V_L +  [I_G] R_G I_G - P_G = 0, \label{eq:f_GV_G}
	\end{split}
	\end{align}
	\begin{align}
	\begin{split}
	\tilde{f}_L (V_L)&=f_L (V_G,V_L)\bigg|_{V_G=V_G^*} ={Y}_{LG} \,V^*_G \, +{Y}_{LL} \,V_L\\&+ Y_LV_L+\bar{I}_L+ \, [V_L]^{-1} \, \bar{P}_L = 0 . \label{eq:f_LV_G}
	\end{split}
	\end{align}
	We will proceed by analyzing equation \eqref{eq:f_LV_G}. Note that $\tilde{f}(V_L^*)=0$ since $V_L^*$ is a feasible solution obtained from the \textbf{SCPF}. Moreover, if the function $\tilde{f}_L(V_L)$ is injective, then $V_L^*$ is the unique solution of \eqref{eq:f_LV_G}. 
	
	To show the injectivity of $\tilde{f}_L(V_L)$, we first evaluate its Jacobian with respect to $V_L$, given as
	\begin{equation}
	\label{eq:Jacobian}
	\JJ(V_L) =\frac{\partial \tilde{f}_L (V_L)}{\partial V_L}={Y}_{LL}+Y_L-\left[[V_L]^{-2}\bar{P}_L\right].
	\end{equation}
	As stated in \cite[Theorem 6]{Gale}, if the Jacobian \eqref{eq:Jacobian} of the function $\tilde{f}_L (V_L)$ is symmetric and positive definite in a convex region $\Omega$, then $\tilde{f}_L (V_L)$ is injective in $\Omega$. Note that $\JJ(V_L)$ is symmetric by construction. 
	Moreover, using Lemma \ref{lemma:yll}, one can split \eqref{eq:Jacobian} into
	\begin{equation}
	\label{eq:Jacobiansplit}
	\JJ(V_L) =\hat{Y}_{LL}+ \underbrace{[-Y_{LG} \textbf{1}_{n}] + Y_L -\left[[V_L]^{-2}\bar{P}_L\right]}_{\widetilde{M}},
	\end{equation}
	where $\hat{Y}_{LL}\succeq 0$ and $-Y_{LG}$ is a nonnegative matrix. For  $\JJ(V_L)$ to be positive definite, it is sufficient to show that $ \tilde{M} \succ 0$. Since $\tilde{M}$ is a diagonal matrix,
	\begin{equation}
	\label{eq:Dpos}
	-  \sum_{j \in \DD}Y_{ij} + Y_{L,i}  - \bar{P}_{L,i}{V^{-2}_{i}} > 0, \quad \forall i \in \tilde{\LL}.
	\end{equation}
	We remark that $-  \sum_{j \in \DD}Y_{ij}$ is positive only if load $i$ is connected directly to at least one DGU, and is otherwise zero. Hence, if 
	\begin{equation}
	\label{eq:Dpos2}
	\bar{P}_{L,i}<{V_{i}^2}\,Y_{L,i},
	\end{equation}
	then \eqref{eq:Dpos} is automatically satisfied and consequently $\JJ (V_L) \succ 0$. Using \eqref{eq:Dpos2}, one can deduce  that  $\tilde{f}_L (V_L)$ is injective in $\Omega$ given as
	\begin{equation*}
	\Omega = \{\,V_{i} : \; V_{i}>\sqrt{\frac{\bar{P}_{L,i}}{{Y}_{L,i}}}, \quad \forall i \in \tilde{\LL} \,\}.
	\end{equation*}
	Since $V_{Li}^* \in [\,V^{min}_{Li}, \, V^{max}_{Li}\,] $ and \eqref{eq:PLcond} holds, $V_L^*$ always belongs to $\Omega$. The uniqueness of $V_L^*$ in $\Omega$ follows from the injectivity of $\tilde{f}_L (V_L)$; moreover, given \eqref{eq:PLcond},  $V_L^*$ is unique in $\YY$. 
	Consequently, considering that $\tilde{f}_G (V^*_L,P_G^*)=0$, it is evident that $P_G =P_G^*$ is the unique solution of \eqref{eq:f_GV_G} if $V_G=V^*_G$ and $V_L=V^*_L$.
\end{proof}
\begin{remark}\textbf{(Condition \eqref{eq:PLcond} and stability).}
	The uniqueness condition \eqref{eq:PLcond} essentially limits the power consumption of P loads. As shown in \cite{Nahata}, due to the negative impedance introduced by the P loads, their power consumption \linebreak$P_{L,i}<(V^*_{i})^2Y_{L,i}, i \in \tilde{\LL}$ in order to guarantee stability. Since $V_i^*$ is the solution of \textbf{SCPF}, $V_i^*\geq V_i^{min}$, by satisfying \eqref{eq:PLcond}, one can simultaneously guarantee the uniqueness of load voltages and the stability of the DCmG.
\end{remark}
%
%
%
\begin{remark}
	The use of a multi-layered hierarchical control scheme is a well-established concept for the overall operation of a mG \cite{Meng}. In the context of islanded DCmGs, supervisory control structures with different functionalities are explored in \cite{Shafiee2014,Kumar, Dragicevic3, Iovine2019}. However, these contributions are restricted to a specific topology, do not consider the interface with the primary layer, or disregard the stability of the DCmG. Besides the incorporation of generic topologies changing over time and the seamless integration of multiple control layers, this work considers both overall mG stability and optimal resource allocation at the same time. Moreover, the secondary control layer can easily be interfaced with any EMS that generates power references. 
\end{remark}
\section{Numerical Results}
\label{sec:simulations}

\begin{figure}[b!]
	\centering
	\ctikzset{bipoles/length=1.2cm}
	\tikzstyle{every node}=[font=\scriptsize]
	\begin{circuitikz}[american currents, scale=0.32]
		
		Defining node styles
		
		\tikzstyle{DGU2} = [circle, double, draw=BurntOrange, fill=BurntOrange!20]
		\tikzstyle{DGU1} = [circle, double, draw=PineGreen, fill=PineGreen!20]
		\tikzstyle{DGU} = [circle, double, draw=Red, fill=Red!20]
		\tikzstyle{loads} = [rectangle,draw=Violet, fill=Violet!20 , text centered, rounded corners, minimum width=0.5cm, double]
		
		 Defining DGU's
		\draw (-1,0) node(6) [DGU2, label={[yshift=-0.05cm,xshift=0.3cm]\textbf{6}}]  {$PV$};
		\draw (-1,6) node(15) [loads, label={[yshift=-0.05cm,xshift=0.3cm]\textbf{15}}]  {$L_c$};
		\draw (-1,10) node(16) [loads, label={[yshift=-0.05cm,xshift=0.3cm]\textbf{16}}]  {$L_c$};
		\draw (3,4) node(1) [DGU, label={[yshift=-0.05cm,xshift=0.3cm]\textbf{1}}]  {$D$};
		\draw (7,0) node(7) [loads, label={[yshift=-0.05cm,xshift=0.3cm]\textbf{7}}]  {$L_a$};
		\draw (7,4) node(14) [loads, label={[yshift=-0.05cm,xshift=0.3cm]\textbf{14}}]  {$L_c$};
		\draw (11,1.5) node(3) [DGU1, label={[yshift=-0.05cm,xshift=0.3cm]\textbf{3}}]  {$B$};
		\draw (11,4) node(4) [DGU1, label={[yshift=-0.05cm,xshift=0.3cm]\textbf{4}}]  {$B$};
		\draw (11,7) node(12) [loads, label={[yshift=-0.05cm,xshift=0.3cm]\textbf{12}}]  {$L_b$};
		\draw (11,10) node(13) [loads, label={[yshift=-0.05cm,xshift=0.3cm]\textbf{13}}]  {$L_c$};
		\draw (15,6) node(2) [DGU, label={[yshift=-0.05cm,xshift=0.3cm]\textbf{2}}]  {$D$};
		\draw (15,0) node(8) [loads, label={[yshift=-0.05cm,xshift=0.3cm]\textbf{8}}]  {$L_a$};
		\draw (19,4) node(11) [loads, label={[yshift=-0.05cm,xshift=0.3cm]\textbf{11}}]  {$L_b$};
		\draw (23,5) node(9) [loads, label={[yshift=-0.05cm,xshift=0.3cm]\textbf{9}}]  {$L_a$};
		\draw (23,10) node(10) [loads, label={[yshift=-0.05cm,xshift=0.3cm]\textbf{10}}]  {$L_b$};
		\draw (23,0) node(5) [DGU1, label={[yshift=-0.05cm,xshift=0.3cm]\textbf{5}}]  {$B$};

		\path [thick] (15) edge [ bend left=10]  (16);	
		\path [thick] (6) edge [ bend left=10]  (15);				
		\path [thick] (15) edge [ bend right=10]  (1);	
		\path [thick] (1) edge [ bend right=10]  (14);	
		\path [thick] (14) edge [ bend right=10]  (4);	
		\path [thick] (4) edge [ bend right=10]  (3);	
		\path [thick] (4) edge [ bend left=10]  (12);	
		\path [thick] (12) edge [ bend left=10]  (13);	
		\path [thick] (6) edge [ bend right=10]  (7);	
		\path [thick] (7) edge [ bend right=10]  (8);	
		\path [thick] (8) edge [ bend right=10]  (5);	
		\path [thick] (5) edge [ bend right=10]  (9);	
		\path [thick] (9) edge [ bend right=10]  (10);	
		\path [thick] (11) edge [ bend right=10]  (9);		
		\path [thick] (12) edge [ bend right=10]  (2);	
		\path [thick] (2) edge [ bend right=10]  (11);	
		
	\end{circuitikz}
	\caption{DCmG based on the modified 16-bus feeder \cite{Low}. The letters $D$, $B$, and $PV$ denote dispatchable, battery, and PV DGUs, respectively. The letter $L$ indicates loads with subscripts $a,b,$ and $c$ defining different consumption patterns.}
	\label{fig:DC_benchmark}
\end{figure}
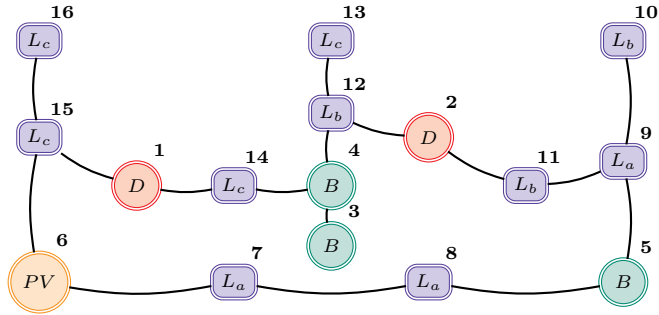

\begin{figure}[!h]
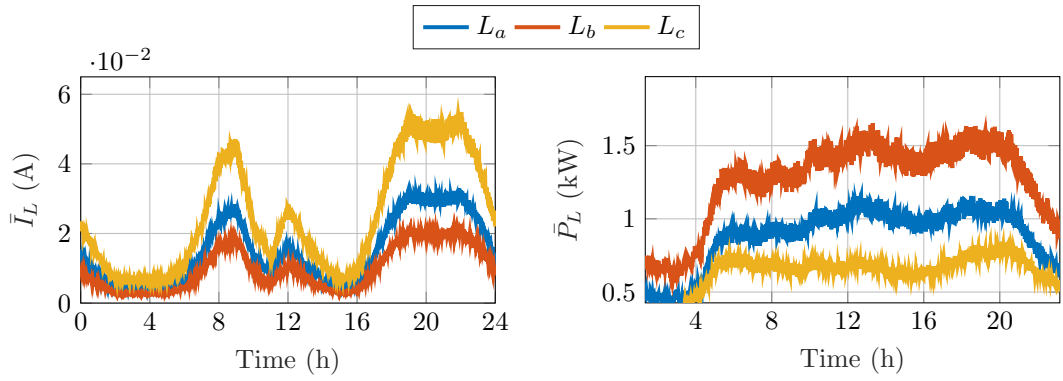

\definecolor{mycolor1}{rgb}{0.00000,0.44700,0.74100}%
\definecolor{mycolor2}{rgb}{0.85000,0.32500,0.09800}%
\definecolor{mycolor3}{rgb}{0.92900,0.69400,0.12500}%
	\setlength\fheight{3cm} 
	\setlength\fwidth{.39\textwidth}
	\centering    
	\ref{currenttrends}\\[-3mm]
	\subfloat{\input{currenttrends.tex}}
	\subfloat{\input{powertrends.tex}}
	\caption{Actual current and power absorption of DCmG loads. Each of the 10 DCmG loads corresponds to one of the three profiles shown above.}
\label{fig:load trends}
\vspace{-5mm}
\end{figure}

In this section, we aim to show the performance of the proposed hierarchical control scheme via simulation studies conducted in MATLAB. We consider the 16-bus DC feeder \cite{Low},equipped with three battery DGUs, two dispatchable DGUs, a PV DGU, and ten ZIP loads (see Figure \ref{fig:DC_benchmark}), in a meshed stand-alone configuration.   The DGUs are interfaced with synchronous Buck converters, and controlled by the primary voltage controllers studied in \cite{Nahata}. We highlight that turning OFF dispatchable DGUs at nodes 1 and 2 simultaneously splits the mG into two separate DCmGs (see Figure \ref{fig:DC_benchmark}). Such an occurrence can be circumvented by adding the simple constraint  $\delta_{D,1}(k+i)+\delta_{D,2}(k+i)\geq 1, \forall i \in [0,\dots,N-1], $ to the EMS optimization problem \eqref{eq:EMSsPF}. The loads are standard ZIP, and their power and current absorption follow three different daily profiles denoted by subscripts $a$, $b$, and $c$ (depicted in Figure \ref{fig:load trends}). The DCmG is operated at a nominal voltage $V^o=100$ Volts, with nodal voltages lying between $V^{min}=0.9V^o$ and $V^{max}=1.1V^o$. The DGU parameters utilized are given in Table \ref{Table1}, \PNREV{while the weights of the EMS cost function \eqref{muG_costfunction} are reported in Table \ref{Table2}.}

{
	\begin{table}[!t]
		\centering
		\begin{tabular}{c c c c c}
			\toprule
			& & & &\\[-1em]
			DGU & $({P}^{\,min},\;{P}^{\,max})$ & $(\eta_{CH},\eta_{DH})$ & $({S}_B^{\,min},\;{S}_B^{\,max}) $ & ${S}_B^o $ 	\\	\midrule 
			D1 & {$(+10, +80)$} & $-$& $-$& $-$ \\
			D2 & {$(+10, +80)$} & $-$& $-$& $-$ \\
			B3 & {$(-40, +40)$} & $(0.9, \,0.9)$ & $(0.1, \,0.9)$  & $0.5$ \\ 
			B4 & {$(-50, +50)$} & $(0.9, \,0.9)$& $(0.1, \,0.9)$ & $0.6$ \\   
			B5 & {$(-60, +60)$} & $(0.9,\, 0.9)$ & $(0.1, \,0.9)$  & $0.4$ \\    
			\bottomrule
		\end{tabular}
		\caption{DGU parameters used by the EMS.}
		\label{Table1}
\end{table}}

\ALBREV{
	\begin{table}[!t]
		\centering
		\begin{tabular}{c c c c c}
			\toprule
			& &\\[-1em]
			DGU	& $w_{D,b}$ & $w_{\delta_D,b}$	\\	\midrule 
			D1 & $254.8$ & $1e6$  \\ 
			D2 & $237.9$ & $1e6$  \\    
			\midrule
			\hspace*{-1em}	& & &\\[-1em]
			&  $w_{B,b}$ & $w_{\delta_B,b}$ & $w_{S,b}$ 	\\	\midrule 
			B3 & {$0.1$} & $5e7$ & $2.5e9$  \\ 
			B4 & {$0.1$} & $5e7$& $2.5e9$  \\   
			B5 & {$0.1$} & $5e7$ & $2.5e9$  \\    
			\midrule
			\hspace*{-1em}	& &\\[-1em]
			& $w_{PV,b}$ & $w_{\delta_{PV},b}$	\\	\midrule 
			PV6 & {$5e4$} & $1e8$  \\ 
			\bottomrule
		\end{tabular}
		\caption{\PNREV{Weights of the EMS cost function \eqref{muG_costfunction}.}}
		\label{Table2}
\end{table}}

{The MPC-based EMS schedules the optimal power set-points for DGUs every 15 minutes, using a prediction horizon of 5 hours, i.e. $N=20$.} The loads in the DCmG network change every minute. With the goal of tracking the received power references despite load variations, the secondary layer runs with a sampling time of 3 minutes. 

In the ensuing discussion, we describe the behaviour of various mG components controlled by the proposed hierarchical controller over a span of 24 hours.

\begin{figure}[!h]
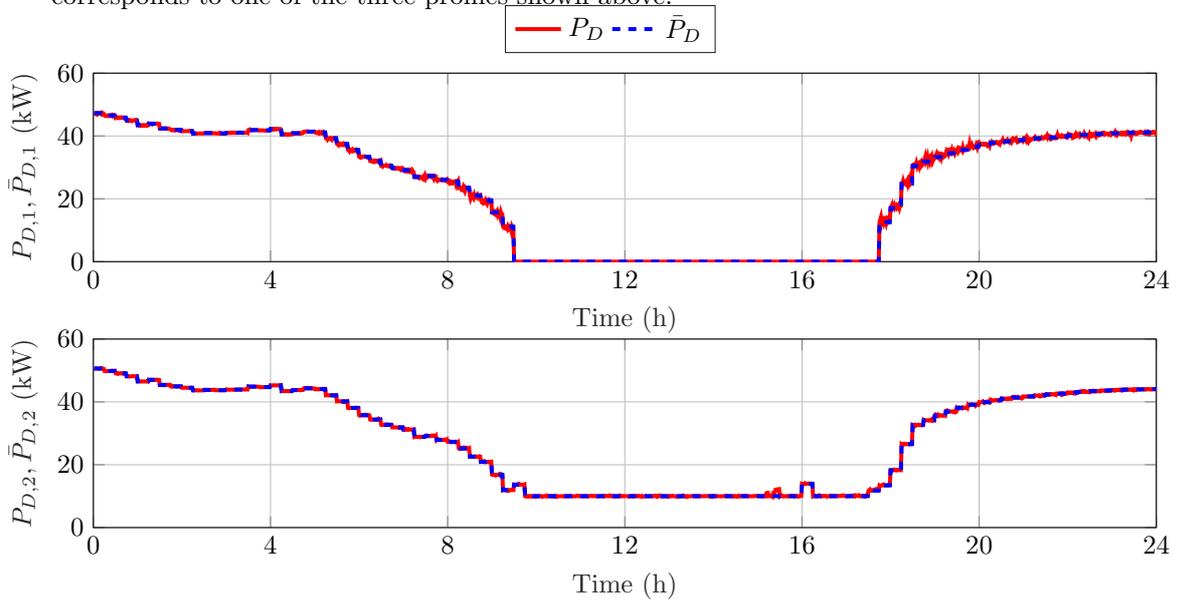

	\setlength\fheight{2.5cm} 
	\setlength\fwidth{1\textwidth}
	\centering
	\ref{dispatchable}
	\subfloat{\hspace{-1.8cm}\input{dispatch1.tex}}\\[-3mm]
	\subfloat{\hspace{-1.8cm}\input{dispatch2.tex}}\\[-3mm]
	\caption{ Power generated by dispatchable DGUs).
	}
	\label{fig:dispatchable}
\end{figure}

\begin{figure}[!h]
	\definecolor{mycolor1}{rgb}{0.00000,0.44700,0.74100}%
	\setlength\fheight{2.5cm} 
	\setlength\fwidth{\textwidth}
	\centering    	
	\ref{PVgenerationforecast} 	\ref{PVgeneration}
	\subfloat{\hspace{-1.9cm}	
%
%
\definecolor{mycolor1}{rgb}{0.00000,0.44700,0.74100}%
\begin{tikzpicture}

\begin{axis}[%
width=0.951\fwidth,
height=\fheight,
at={(0\fwidth,0\fheight)},
scale only axis,
xmin=0,
xmax=24,
xtick={ 0,  4,  8, 12, 16, 20, 24},
xlabel style={font=\color{white!15!black}},
xlabel={Time (h)},
ymin=-20,
ymax=160,
ylabel style={font=\color{white!15!black}},
ylabel={Power (kW)},
axis background/.style={fill=white},
xmajorgrids,
ymajorgrids,
legend columns=-1,
legend entries={${P}^o_{PV,6}$, ${P}^f_{PV,6}$},
legend to name=PVgenerationforecast,
legend style={at={(0.5,0.97)}, anchor=north, legend columns=1, legend cell align=left, align=left, draw=white!15!black}
]

\addplot [color=mycolor1, dotted, line width=1.5pt]
  table[row sep=crcr]{%
0.0166666666666799	-0\\
4.01666666666668	-0\\
4.26666666666668	1.29937018258582\\
4.51666666666668	2.6212364912339\\
4.76666666666668	3.9037678301122\\
5.01666666666668	5.20054562783545\\
6.01666666666668	26.0734356646702\\
6.26666666666668	29.9494651653621\\
6.51666666666668	33.7784827006792\\
6.76666666666668	37.8674834091034\\
7.01666666666668	41.6809574935536\\
7.26666666666668	46.8788190618963\\
7.51666666666668	52.3553380273181\\
7.76666666666668	57.3044906340345\\
8.01666666666668	62.7884504726479\\
8.26666666666668	67.935021333307\\
8.51666666666668	73.4600737148197\\
8.76666666666668	78.4754198519462\\
9.01666666666668	83.7558833421705\\
9.51666666666668	96.9182567676629\\
9.76666666666668	103.15953131887\\
10.0166666666667	109.995566629237\\
10.2666666666667	112.528768388122\\
10.5166666666667	114.439584230143\\
10.7666666666667	117.070474758389\\
11.0166666666667	119.746064986483\\
11.2666666666667	124.30404937892\\
11.5166666666667	127.714035392196\\
11.7666666666667	132.483554642134\\
12.0166666666667	135.320819908836\\
12.2666666666667	137.602942935566\\
12.5166666666667	137.766704654892\\
12.7666666666667	140.017387595924\\
13.0166666666667	140.752599403687\\
13.2666666666667	142.545910286132\\
13.5166666666667	143.000009136764\\
13.7666666666667	144.874182871372\\
14.0166666666667	146.103715642044\\
14.2666666666667	146.454032813802\\
14.5166666666667	146.247334382518\\
14.7666666666667	146.283017326924\\
15.0166666666667	141.763207262283\\
15.2666666666667	118.403245955834\\
15.5166666666667	107.721050630983\\
15.7666666666667	120.039355623493\\
16.0166666666667	135.821683138441\\
16.2666666666667	130.864688132067\\
16.5166666666667	125.485101950174\\
16.7666666666667	119.716954249217\\
17.0166666666667	114.352934064578\\
17.2666666666667	105.498678909598\\
17.5166666666667	96.8704249330453\\
17.7666666666667	87.8478499110262\\
18.0166666666667	78.3875929231752\\
18.2666666666667	66.7614770078571\\
18.5166666666667	54.6819477095256\\
18.7666666666667	42.9967469916207\\
19.0166666666667	31.4519472721804\\
20.0166666666667	-0\\
23.7666666666667	-0\\
};

\addplot [color=black, dashed, line width=1.5pt]
  table[row sep=crcr]{%
0.25	-0\\
4.75	-0\\
5	1.64198775420201\\
5.25	8.26426830313113\\
5.5	14.7456770826877\\
5.75	21.0834892596379\\
6	27.2749854921669\\
6.25	33.3174728890921\\
6.5	39.2083067293368\\
6.75	44.9449179907996\\
7	50.5248444878137\\
7.25	55.9457668553495\\
7.5	61.2055500118515\\
7.75	66.3022861562305\\
8	71.234344273382\\
8.25	76.00043142293\\
8.5	80.5996714247225\\
8.75	85.0316970027312\\
9	89.29676331069\\
9.25	93.3958895896557\\
9.5	97.3310296587644\\
9.75	101.105267603772\\
10	104.723060268167\\
10.25	108.190554243896\\
10.5	111.51600591034\\
10.75	114.710360710399\\
11	117.787849751968\\
11.25	120.765928095402\\
11.5	123.662334399211\\
11.75	126.482891526575\\
12	129.188565204002\\
12.25	131.659877213666\\
12.5	133.691711213734\\
12.75	135.310567286366\\
13	136.448297440252\\
13.25	137.104339740749\\
13.5	137.301244046695\\
13.75	137.061970854909\\
14	136.404352109368\\
14.25	135.34112230726\\
14.5	133.88105850646\\
14.75	132.030075669908\\
15	129.79202168446\\
15.25	127.169251210588\\
15.5	124.163109151144\\
15.75	120.774421363674\\
16	117.003749989725\\
16.25	112.851328317391\\
16.5	108.316875812707\\
16.75	103.399666545175\\
17	98.0986365907402\\
17.25	92.4124666142329\\
17.5	86.3396481333421\\
17.75	79.8785481267792\\
18	73.0274611930027\\
18.25	65.7846474550474\\
18.5	58.1483665118782\\
18.75	50.1169190998206\\
19	41.6886934840722\\
19.25	32.8622163532191\\
19.5	23.6362039649202\\
19.75	14.0095828689519\\
20	3.98149294311287\\
20.25	-0\\
24	-0\\
};

\end{axis}

\begin{axis}[%
width=1.227\fwidth,
height=1.227\fheight,
at={(-0.16\fwidth,-0.135\fheight)},
scale only axis,
xmin=0,
xmax=1,
ymin=0,
ymax=1,
axis line style={draw=none},
ticks=none,
axis x line*=bottom,
axis y line*=left,
legend style={legend cell align=left, align=left, draw=white!15!black}
]
\end{axis}
\end{tikzpicture}%
}\\[-3mm]
	\centering    	
	\subfloat{\hspace{-1.9cm}	\input{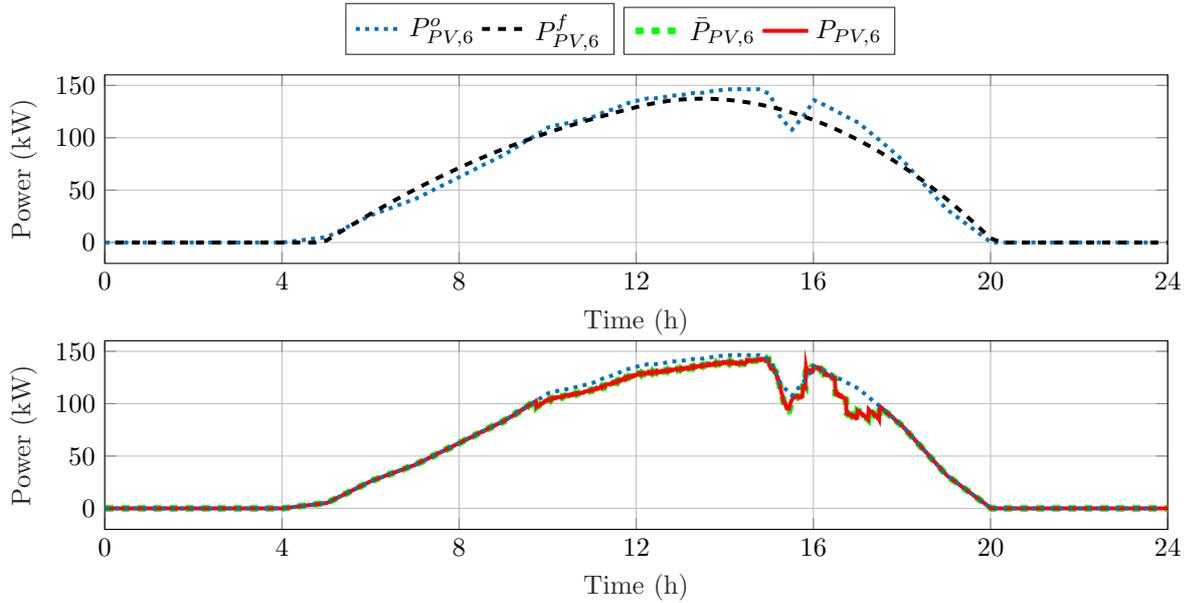}}\\[-3mm]
	\caption{Nominal generation, PV generation forecast, EMS power reference,  generated power for DGU PV6.   
	}
\label{fig:PVDGU}
\vspace{-0.1cm}
\end{figure}

\textbf{Dispatchable DGUs:} As shown in Figure \ref{fig:dispatchable}, DGUs D1 and D2 follow the power references provided by the EMS. During the day, when PV generation starts picking up (see Figure \ref{fig:PVDGU}), the EMS turns OFF DGU D1 to ensure economic optimality and maintain mG power balance. DGU D2, although producing minimum permissible power during the period of peak PV generation, remains operational throughout the day in order to maintain connectivity of the DCmG.%
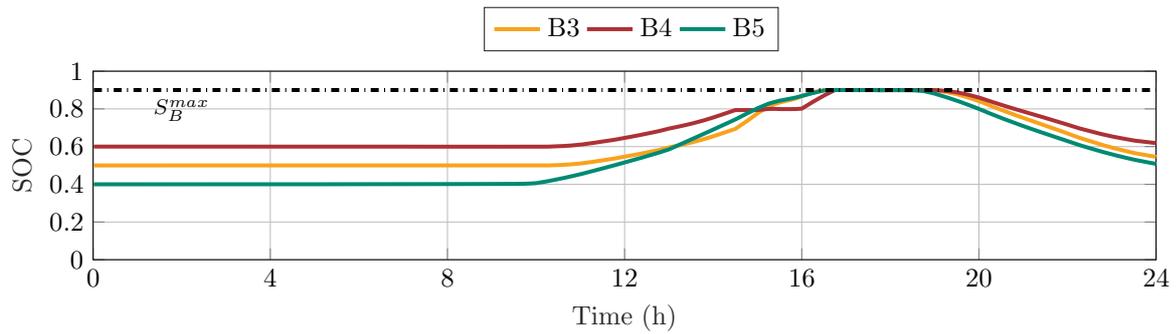
\begin{figure}[!h]
	\definecolor{mycolor1}{rgb}{0.00000,0.44700,0.74100}%
	\definecolor{mycolor2}{rgb}{0.85000,0.32500,0.09800}%
	\definecolor{mycolor3}{rgb}{0.92900,0.69400,0.12500}%
	\setlength\fheight{2.5cm} 
	\setlength\fwidth{\textwidth}
	\centering    	
	\ref{BatteryS0C}
	\subfloat{\hspace{-2.2cm}	
%
%
\definecolor{mycolor1}{rgb}{0.00000,0.44700,0.74100}%
\definecolor{mycolor2}{rgb}{0.85000,0.32500,0.09800}%
\definecolor{mycolor3}{rgb}{0.92900,0.69400,0.12500}%
\begin{tikzpicture}

\begin{axis}[%
width=0.951\fwidth,
height=\fheight,
at={(0\fwidth,0\fheight)},
scale only axis,
xmin=0,
xmax=24,
xtick={ 0,  4,  8, 12, 16, 20, 24},
xlabel style={font=\color{white!15!black}},
xlabel={Time (h)},
ymin=0,
ymax=1,
axis background/.style={fill=white},
xmajorgrids,
ymajorgrids,
ylabel style={font=\color{white!15!black}},
ylabel={SOC},
legend columns=-1,
legend entries={${P}_{B}$, $\bar{P}_{B}$},
legend to name=BatteryS0C,
legend style={at={(0.5,0.97)}, anchor=north, legend columns=1, legend cell align=left, align=left, draw=white!15!black}
]

\addplot [color=YellowOrange, line width=1.5pt]
  table[row sep=crcr]{%
0.0166666666666657	0.5\\
10.2833333333333	0.500255401447067\\
10.55	0.502485722980055\\
10.7833333333333	0.505747035170053\\
11.0166666666667	0.510723836282466\\
11.3	0.519874347462984\\
11.6333333333333	0.531800154834016\\
11.7833333333333	0.537262020534314\\
12.3166666666667	0.560279383227506\\
12.6333333333333	0.575285461441613\\
12.7666666666667	0.581936094091557\\
13.0166666666667	0.59651914830329\\
13.3	0.61047143121268\\
13.5166666666667	0.622187382211717\\
13.7666666666667	0.637916946390998\\
14.0333333333333	0.657114013068664\\
14.5	0.693901149707727\\
14.7666666666667	0.736058914692045\\
15	0.77634599159197\\
15.25	0.812556773831098\\
15.3833333333333	0.824662567027104\\
15.4333333333333	0.828763788260957\\
15.5	0.834036859595486\\
15.7666666666667	0.848847595591167\\
15.85	0.85420108864442\\
16.0833333333333	0.873704846789401\\
16.25	0.888501848609266\\
16.5	0.90000827268576\\
17.6166666666667	0.899784090481617\\
17.85	0.899377651677085\\
18.95	0.89911095389995\\
19	0.899046459359838\\
19.2666666666667	0.889412574976369\\
19.5	0.87787052694004\\
19.7666666666667	0.859816452994117\\
20.0166666666667	0.839833398307356\\
20.3166666666667	0.811686411414939\\
20.5833333333333	0.787710538309828\\
20.8166666666667	0.767548036053661\\
21.1666666666667	0.738637339693724\\
21.8333333333333	0.683900048791198\\
22.15	0.656819647491204\\
22.2666666666667	0.646940722269076\\
22.8	0.607809355949126\\
23.0333333333333	0.592713413874275\\
23.4	0.571679946210566\\
23.5166666666667	0.565350363971248\\
23.8166666666667	0.552952964140871\\
24	0.545967682917762\\
};
\addlegendentry{B3}
\addplot [color=Maroon, line width=1.5pt]
table[row sep=crcr]{%
	0.0166666666666657	0.600000000000001\\
	7.08333333333333	0.599753198979627\\
	7.31666666666667	0.599635860801133\\
	7.88333333333333	0.599551951489083\\
	8.01666666666667	0.599518335135944\\
	8.68333333333333	0.599520270001264\\
	8.83333333333334	0.599511237500558\\
	9.28333333333333	0.599464877429888\\
	9.48333333333333	0.599587408886304\\
	9.6	0.599593880546973\\
	10.2666666666667	0.599720647377517\\
	10.6333333333333	0.603337768728409\\
	10.7833333333333	0.605555898443143\\
	11.05	0.61160288917565\\
	11.3166666666667	0.620246714821594\\
	11.6166666666667	0.630867095908776\\
	11.7666666666667	0.63610687814943\\
	12.0333333333333	0.647485574726705\\
	12.15	0.652472656657896\\
	12.3	0.659164551921222\\
	12.5666666666667	0.671832982950576\\
	12.7666666666667	0.68177519376729\\
	13.0166666666667	0.696376371545561\\
	13.15	0.702772487055579\\
	13.3	0.710301073577259\\
	13.5333333333333	0.723039094181413\\
	13.7833333333333	0.738843994866823\\
	14.0166666666667	0.755591219223529\\
	14.5	0.793713490374717\\
	15	0.793831165094662\\
	15.2333333333333	0.799369025325404\\
	15.2666666666667	0.799757590648241\\
	15.3833333333333	0.800082617461005\\
	15.4333333333333	0.799717561103286\\
	15.5333333333333	0.798957757202057\\
	15.7166666666667	0.798902462525611\\
	15.8666666666667	0.799154930183292\\
	16	0.801318679556502\\
	16.2666666666667	0.839355851671886\\
	16.5	0.870728231637862\\
	16.75	0.899923808784212\\
	17.1166666666667	0.899744103702051\\
	17.3666666666667	0.899761837013248\\
	17.5166666666667	0.899882634082097\\
	17.8166666666667	0.899599580805891\\
	18.0666666666667	0.90002745092556\\
	18.2333333333333	0.899904805278261\\
	18.4333333333333	0.899909077957386\\
	18.6166666666667	0.899972920439708\\
	19.0166666666667	0.899449838757548\\
	19.1	0.898192617530505\\
	19.25	0.895806007336411\\
	19.5	0.88801317888559\\
	19.7666666666667	0.874783771440434\\
	20.0166666666667	0.859016352214102\\
	20.1333333333333	0.84992312172162\\
	20.25	0.840687717384153\\
	20.6	0.814496497793471\\
	20.8	0.800152543715047\\
	21.1833333333333	0.773752494823849\\
	21.3166666666667	0.764650014834277\\
	21.4833333333333	0.753106081679398\\
	22	0.716366870053026\\
	22.2666666666667	0.696734454899378\\
	22.75	0.667209285534586\\
	23.0166666666667	0.653266693909035\\
	23.2666666666667	0.642123909067109\\
	23.5166666666667	0.631649724613741\\
	23.9333333333333	0.6202819108968\\
	24	0.618612414042136\\
};
\addlegendentry{B4}
\addplot [color=PineGreen, line width=1.5pt]
  table[row sep=crcr]{%
0.0166666666666657	0.399999999999999\\
5.21666666666667	0.400287877617394\\
5.85	0.400595516788592\\
6.96666666666667	0.400880750919722\\
7.83333333333333	0.401213624381697\\
8.05	0.401238062246634\\
8.48333333333333	0.401497726774924\\
8.75	0.401665932935281\\
9.08333333333334	0.401934969959939\\
9.2	0.402003328992766\\
9.53333333333333	0.40228805637183\\
9.65	0.402727084305333\\
9.75	0.402942516360127\\
10	0.405908371466836\\
10.2833333333333	0.417199632130004\\
10.6166666666667	0.433032331725485\\
10.8333333333333	0.444203266742232\\
11.05	0.456293339872335\\
11.35	0.474733784903474\\
11.6333333333333	0.492395423131065\\
11.8	0.502762890404913\\
12.3666666666667	0.540021520652768\\
12.7333333333333	0.565125539025356\\
12.8	0.570021878009673\\
13	0.585113054475865\\
13.4166666666667	0.628732872757681\\
13.65	0.653203747881321\\
13.9333333333333	0.683057961877523\\
14.15	0.706368302905506\\
14.4166666666667	0.73497093068007\\
14.5166666666667	0.745899095107514\\
14.7833333333333	0.778091165824886\\
15	0.803118745790226\\
15.25	0.827767150316991\\
15.4	0.837610995732391\\
15.5	0.843550069731648\\
15.8	0.857099529838994\\
15.8666666666667	0.860604699115846\\
16.3	0.886792786333206\\
16.5	0.897311879004896\\
16.7666666666667	0.899963383118553\\
17.6166666666667	0.899625432336236\\
17.7833333333333	0.898927573430925\\
17.9333333333333	0.899368659446981\\
18.1166666666667	0.899490093768623\\
18.4833333333333	0.898998125523487\\
18.5166666666667	0.898685236679643\\
18.75	0.894761966026451\\
18.9333333333333	0.884733851827775\\
19.0166666666667	0.87982655936236\\
19.3333333333333	0.856664050114176\\
19.5666666666667	0.838128171079099\\
19.8166666666667	0.816433788648592\\
20.0166666666667	0.798195009550945\\
20.25	0.775556679251132\\
20.5666666666667	0.746373161495523\\
20.9833333333333	0.710328650054127\\
21.2	0.692497594814267\\
21.3833333333333	0.677727735761735\\
21.4833333333333	0.669785011237543\\
21.6666666666667	0.655317501853908\\
22.2833333333333	0.607047113923766\\
22.4166666666667	0.597719328067722\\
22.6333333333333	0.582692881109352\\
22.8166666666667	0.570452032742633\\
23.1	0.553011892910622\\
23.3166666666667	0.540722551451676\\
23.5333333333333	0.529045561534559\\
23.8666666666667	0.514126177775854\\
24	0.508536938548879\\
};
\addlegendentry{B5}

\addplot [color=black, dashdotted, line width=1.5pt]
table[row sep=crcr]{%
	0.0166666666666657	0.9\\
	24	0.9\\
};
\node[black, below] at (axis cs:2,0.9){\footnotesize{$S_B^{max}$}};

\end{axis}

\begin{axis}[%
width=1.227\fwidth,
height=1.227\fheight,
at={(-0.16\fwidth,-0.135\fheight)},
scale only axis,
xmin=0,
xmax=1,
ymin=0,
ymax=1,
axis line style={draw=none},
ticks=none,
axis x line*=bottom,
axis y line*=left,
legend style={legend cell align=left, align=left, draw=white!15!black}
]
\end{axis}
\end{tikzpicture}%
}\\[-3mm]
	\caption{ States of charge of DGUs B3, B4, and B5.
	}
	\label{fig:Battery_SOC}
\end{figure}

\begin{figure}[!h]
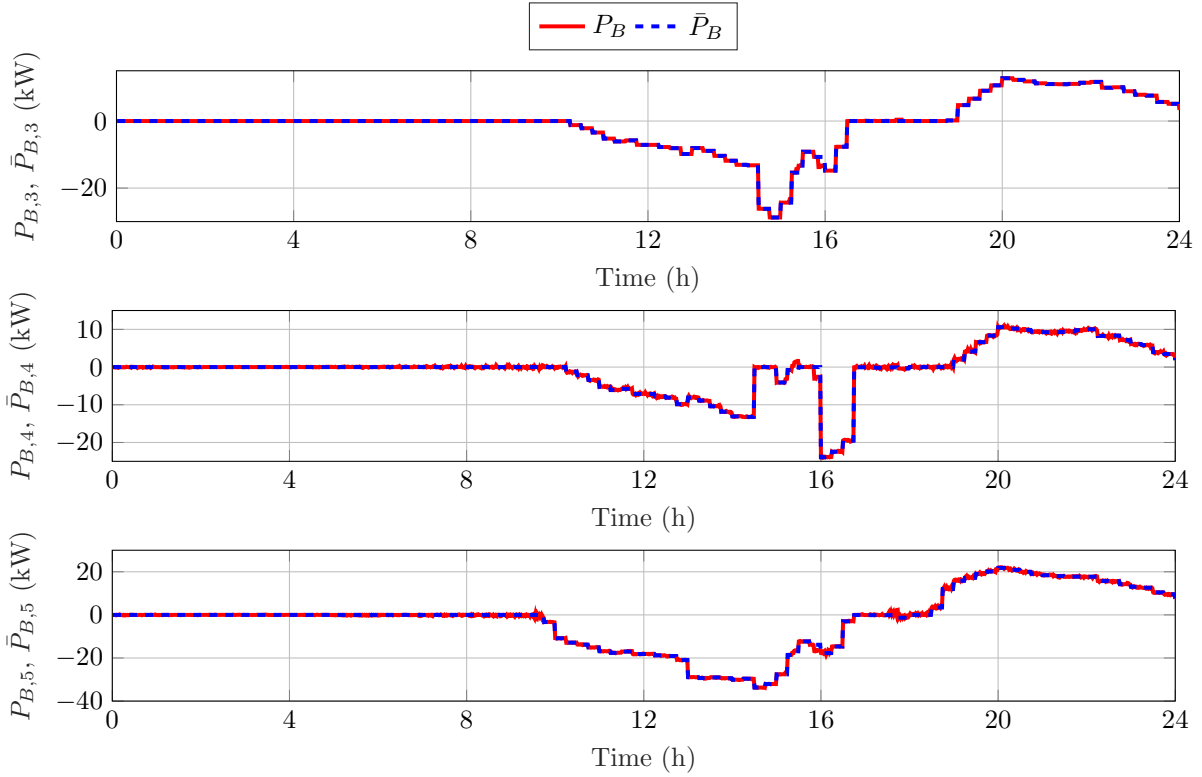

	\setlength\fheight{2cm} 
	\setlength\fwidth{1\textwidth}
	\centering
	\ref{Battery}\\[-1.5mm]
	\subfloat{\hspace{-1.8cm}\input{Battery2.tex}}\\[-3mm]
	\subfloat{\hspace{-1cm}\input{Battery1.tex}}\\[-3mm]
	\subfloat{\hspace{-1cm}\input{Battery3.tex}}\\[-3mm]
	\caption{Power output by battery DGUs.
	}
	\label{fig:Battery_outputs}
\end{figure}

\textbf{Battery DGUs:} In Figure \ref{fig:Battery_outputs}, notice that battery DGUs follow power references provided by the EMS. Working to the detriment of battery's longevity, abrupt charging and discharging, and frequent switching between these two modes are prevented by the EMS. As for the SOCs---reported in Figure \ref{fig:Battery_SOC}, they evolve while respecting the operational constraints.  Moreover, the EMS tries to store surplus energy during periods of peak PV generation (see Figure \ref{fig:PVDGU}). This energy is released later in the day when the PV generation declines.	

\textbf{PV DGU:} We conducted the simulations with a mismatch between nominal PV generation and forecasts, so as to be consistent with a real operation scenario (refer to Figure \ref{fig:PVDGU}).  At each sampling instant, the EMS utilizes the nominal PV generation and the forecast not only to generate power references but also to decide whether to operate the PV DGU in MPPT or power curtailment mode.  As seen from Figure \ref{fig:PVDGU}, the power injected into the DCmG closely tracks the EMS power references. Notice that the PV DGU operates in MPPT mode during the first and the last hours of the simulation, whereas it curtails power during the central part of the day. \PNREV{A power curtailment is clearly inevitable at around 15h considering that (i) the SOCs are about to hit their upper bound, (ii) DGU D1 is nonoperational, and (iii) DGU D2---cannot be switched OFF---is injecting minimum power.}
 
\begin{figure}[!h]
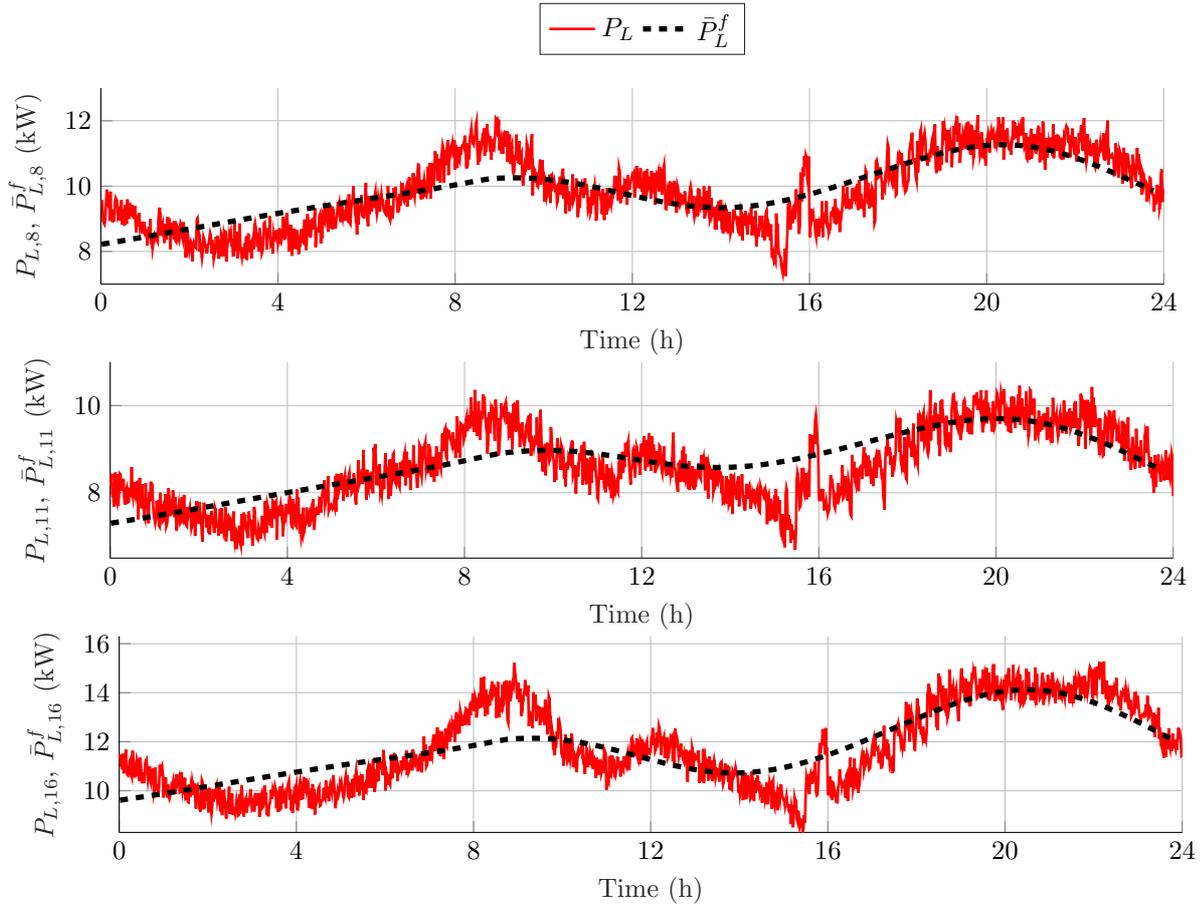

	\definecolor{mycolor1}{rgb}{0.00000,0.44700,0.74100}%
	\definecolor{mycolor2}{rgb}{0.85000,0.32500,0.09800}%
	\definecolor{mycolor3}{rgb}{0.92900,0.69400,0.12500}%
	\setlength\fheight{2.6cm} 
	\setlength\fwidth{1\textwidth}
	\centering
	\ref{load}\\
	\subfloat{\hspace{-1.3cm}\input{load3.tex}}\\[-3mm]
	\subfloat{\hspace{-2cm}\input{load2.tex}}\\[-3mm]
	\subfloat{\hspace{-2cm}	\input{load1.tex}}\\[-3mm]
	\caption{Load power forecasts and net power absorption for different load nodes.
	}
	\label{fig:load}
\end{figure}

\textbf{Loads:} The load power forecasts used by the EMS and the net power absorption for nodes 8, 11, and 16 are shown in Figure \ref{fig:load}. One can observe that the forecasts are fairly different from the actual power absorption. This stems from the fact that EMS forecasts are deduced  at nominal voltage through inaccurate current and power profiles (see Figure \ref{fig:load trends} for actual current and power absorption). Even if exact profiles were available to the EMS \textit{a priori}, the forecasts would not coincide with net power absorbed by the loads. This is because the net power absorbed by a load depends on \PNREV{PC} voltages---generated by the secondary layer only after EMS power references are received.
\begin{figure}[h!]
	\centering
	\setlength\fheight{3cm} 
	\setlength\fwidth{1\textwidth}
	\subfloat{\hspace{-2cm}	\input{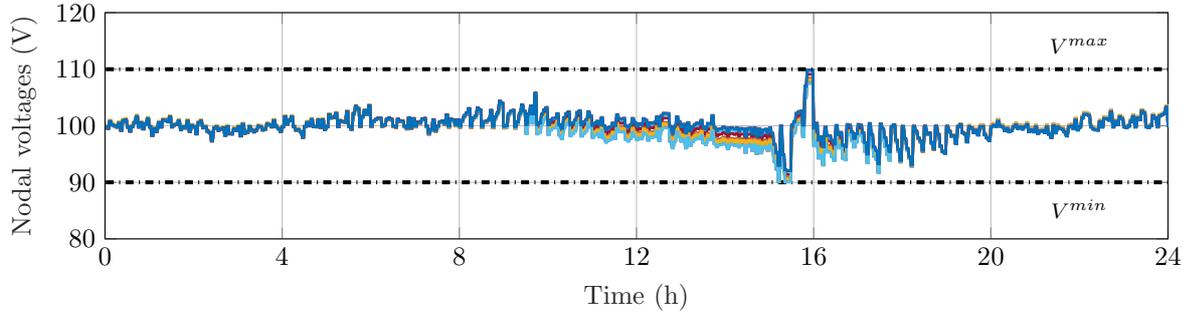}}\\[-3mm]
	\caption{Nodal voltages in the DCmG network.}
	\label{fig:voltages}
\end{figure}
\PNREV{As the loads change over the course of a secondary sampling interval, one can observe small red spikes around the EMS power references in Figures \ref{fig:dispatchable},\ref{fig:Battery_outputs}, and \ref{fig:PVDGU}. The DGU power $P_G$ is restored to the reference value at the next secondary sampling time.}

\begin{figure}[!h]
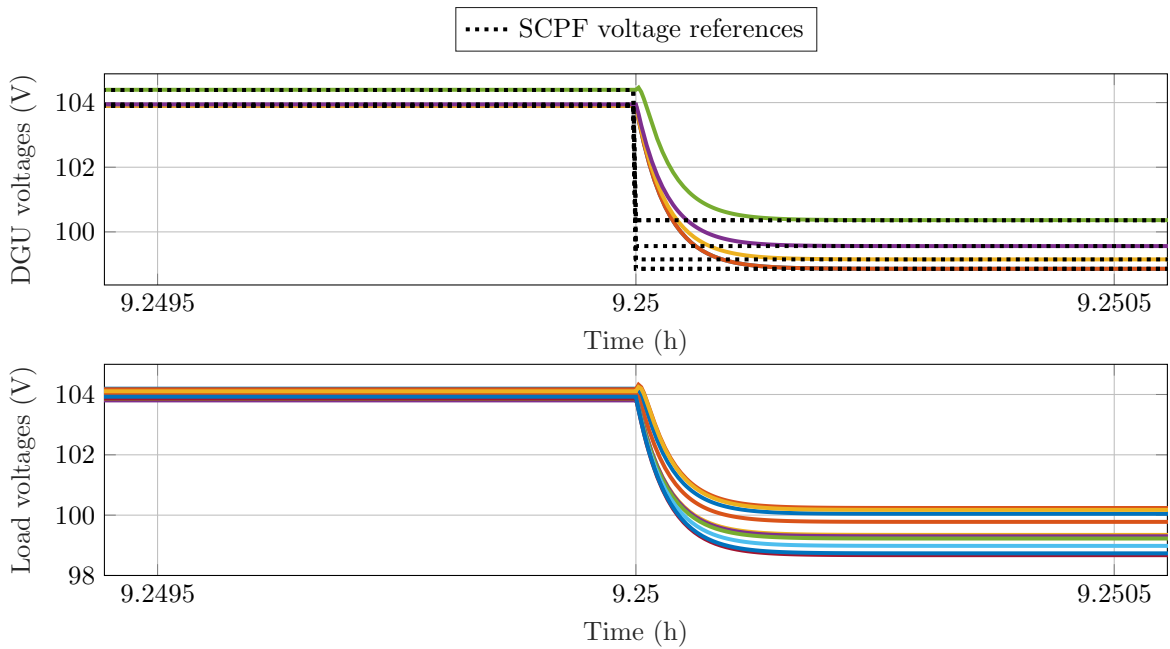

	\setlength\fheight{2.8cm} 
	\setlength\fwidth{1\textwidth}
	\centering
	\ref{DGUplugout}
    \subfloat{\hspace{-2cm}\input{PlugoutDGUs.tex}}\\[-3mm]
	\subfloat{\hspace{-2cm}\input{DGUgenplugoutloads.tex}}\\[-3mm]
	\caption{Nodal voltages when dispatchable  DGU D1 is turned off.
	}
	\label{fig:Gen_voltages}
\end{figure}

Finally, we highlight that, during the simulation, condition \eqref{eq:PLcond} always holds for all load nodes, ensuring the uniqueness of solution for load voltages. The secondary control layer maintains the voltages in the allowed range, as shown in Figure \ref{fig:voltages}.  As a consequence of new power references received from the EMS, a clear change in voltages can be observed every 15 minutes. In Figure \ref{fig:Gen_voltages}, we show the performance of primary voltage controllers when dispatchable DGU D1 is turned OFF by the EMS. Indeed, thanks to the implemented plug-and-play primary controllers, the transients quickly die out and voltages are forced back to desired references.
\PNREV{
\begin{remark}\textbf{(Practical implementation aspects of our hierarchical control scheme).}
The supervisory controller comprising primary and secondary control layers can be implemented in a single central unit. On the contrary, each DGU is equipped with a primary voltage regulator which along with other such regulators constitute the primary control layer. The secondary layer requires measurements of loads consumption, and assumes knowledge of the DCmG admittance matrix $Y$. The tertiary layer, on the other hand, does not hinge on $Y$, but requires additional measurements like battery SOC and nominal PV power. 
\\
\\
We note that high computational performance is not required to execute our supervisory controller. The presented 16-bus test-case has been simulated on a personal computer with an Intel Core i7-6500u processor. The tertiary EMS layer computed the optimal solution in $2s$ on an average, while the secondary layer averaged at $1s$ to perform the power-to-voltage translation. Acting on the order of few microseconds, primary voltage control schemes \cite{Martinelli2018, Nahata, strehle2020scalable, cucuzzella2017decentralized} can easily be utilized with our supervisory control layer. 
\end{remark}
}

\section{Conclusions}
\label{sec:conclusion}
\PNREV{In this work, we proposed a top-to-bottom hierarchical control structure for an islanded DCmG. Our supervisory controller resting atop a primary voltage layer comprises secondary and tertiary layers. By putting to use load/generation forecasts, as well as measurements drawn from loads and DGUs, our EMS--equipped tertiary layer generates optimal power references. To render EMS power references meaningful for the voltage-controlled primary layer, the secondary layer translates them into apposite voltage references. More specifically, the voltage references are generated by virtue of an optimization problem capable of incorporating practical operational constraints like DGU capability limits and DCmG permissible voltage range. We studied the well-possessedness of the secondary optimization problem, and deduced a novel condition for the uniqueness of generator voltages and DGU power injections. Lastly, on a 16-node DCmG, we demonstrated how multiple layers of our hierarchical control scheme work in tandem to achieve desired objectives. Future work will target the development of an EMS which enables the DCmG to work in grid-connected mode and a secondary layer that do not need a complete knowledge of the DCmG topology. With the aim of designing a fully scalable DCmG control architecture, efforts will be also directed in the direction of decentralization of secondary and tertiary control layers.}
\vspace*{-3mm}

\bibliographystyle{IEEEtran}
\bibliography{articleref}
\setcounter{section}{0}
\renewcommand*{\thesection}{\Alph{section}}
\section{Appendix} 
\smallskip
\begin{lemma}
	\label{lemma:yll}
	The  matrix ${Y}_{LL}$ can be written as \begin{equation}\label{eq:lapsub} {Y}_{LL}=\hat{Y}_{LL}+ [-Y_{LG} \textbf{1}_{n}],\end{equation} where $\hat{Y}_{LL}$ is a Laplacian matrix .
\end{lemma}
\begin{proof}
	The network admittance matrix ${Y}$ is a Laplacian with zero row sum \cite{dorfler2018electrical}. Matrix $Y_{{LL}}$, a submatrix of $Y$, is symmetric with positive diagonal and non-negative off-diagonal entries. Since the network graph $\GG$ is connected, $Y_{{LL}}$ has at least one row with strictly positive row sum.  $Y_{LL}$ is a Laplacian matrix with self loops \cite{dorfler2013kron} and, therefore, can be written as \eqref{eq:lapsub}.
\end{proof}
\begin{lemma}
	\label{lemma:yinv_ylg}
	The matrix $-({Y}_{LL}+Y_L)^{-1} \,{Y}_{LG}$ has no rows with all zero entries and is nonnegative.
\end{lemma}
\begin{proof}
	The matrix $-{Y}_{LG}$ is a non-negative matrix and, since the graph is connected, has at least one row with non-zero row sum. 
	The statement of the above Lemma follows from the fact that ${Y}_{LL}+Y_L$ is a Laplacian matrix with self loops, the inverse of which is strictly positive \cite{dorfler2013kron}.
\end{proof}
\subsection{Proof Of Proposition }
	Under Assumption \PNREV{\ref{ass:Vpos}}, equations \eqref{eq:CPF_G} and \eqref{eq:CPF_L} can be expressed in a single matrix equality as follows
	\begin{align}
	\begin{split}
	f(V,P_G) &= [V]\, \tilde{Y} \, V \;+\;   [V] \widetilde{I} \;+ 
	\begin{bmatrix}
	\,[I_G] \, R \, I_G \, \\ \mbf{0}
	\end{bmatrix}  \\&+ 
	\begin{bmatrix}
	\, -P_G \, \\ \bar{P}_L
	\end{bmatrix}  =   \mbf{0}_{n+m},
	\label{f_tot}
	\end{split}
	\end{align}
	where $ \widetilde{I} = \begin{bmatrix}
	\mbf{0}^T_n &  \bar{I}_L^{\,T} \,
	\end{bmatrix}^T$, and $\tilde{Y}= Y + \begin{bmatrix}
	\mbf{0} &\mbf{0}\\
	\mbf{0} &Y_L
	\end{bmatrix}.$
	To achieve $J^*_{\scriptscriptstyle SPF} =0$, a solution $(V,P_G)$ to \textbf{SPF} must exist such that $P_G=\bar{P}_G$ and
	\begin{equation}
	f(V,\bar{P}_G)=\mbf{0}_{n+m}.
	\label{eq:f_tot_barp}
	\end{equation}
	On multiplying the above equation by $\mbf{1}_{n+m}^T$ on both sides, one obtains 
	\begin{align}
	\begin{split}
	\mbf{1}_{n+m}^T \, f(V,\bar{P}_G) &=  \,V^T\, \tilde{Y} \, V \,+  \, V^T \, \widetilde{I}\,+  \,
	I_G^T \, R \, I_G \,\\& - \mbf{1}_{n}^T\, P_G \, + \,\mbf{1}_{m}^T\, \bar{P}_L
	\, =  \, 0 \,.
	\end{split}
	\label{eq:tf_tot_barp1}
	\end{align}
	Bear in mind that a solution to \eqref{eq:f_tot_barp} also verifies \eqref{eq:tf_tot_barp1}.
\PNREV{Grouping the terms $ \,V^T\, \tilde{Y} \, V $ and $V^T \, \widetilde{I}$ in equation \eqref{eq:tf_tot_barp1} together yields
\begin{align}
\begin{split}
&(V+\frac{1}{2} \tilde{Y}^{-1} \widetilde{I})^T \, \tilde{Y} \, (V+\frac{1}{2} \tilde{Y}^{-1} \widetilde{I}) +  \,
I_G^T \, R \, I_G \\& - \mbf{1}_{n}^T\, P_G \, + \,\mbf{1}_{m}^T\, \bar{P}_L
\, =  \, 0 \,,
\end{split}
\end{align}
which then becomes
\begin{align}
\begin{split}
&(V+\frac{1}{2} \tilde{Y}^{-1} \widetilde{I})^T \, \tilde{Y} \, (V+\frac{1}{2} \tilde{Y}^{-1} \widetilde{I}) +  \,
I_G^T \, R \, I_G \\&= \; \frac{1}{4} \, \widetilde{I}^T  \,\tilde{Y}^{-1} \, \widetilde{I} + \sum_{\forall i \in \DD} \bar{P}_G \, - \,\sum_{\forall i \in \LL} \bar{P}_L.
\end{split}
\label{eq:balance}
\end{align}
}
Note that the matrices $\tilde{Y} \succ 0$ and $R_G \succ 0$,  and hence, if a voltage solution $V$ exists, then
\begin{equation}
	(V+\frac{1}{2} \tilde{Y}^{-1} \widetilde{I})^T \, \tilde{Y} \, (V+\frac{1}{2} \tilde{Y}^{-1} \widetilde{I}) +  \,
	I_G^T \, R \, I_G \, \geq 0,
	\label{eq:balance1}
	\end{equation}
	\PNREV{Therefore,}
	\begin{equation}
	\frac{1}{4} \, \widetilde{I}^{\;T}  \,\tilde{Y}^{-1} \, \widetilde{I} + \sum_{\forall i \in \DD} \bar{P}_G  \, - \,\sum_{\forall i \in \LL} \bar{P}_L  \geq 0.
	\label{eq:spf_pre_neccond}
	\end{equation}
	\PNREV{Using established results on the inverse of block matrices \cite[Theorem 2.1]{LU}, one can simplify the expression $ \widetilde{I}^{\;T}  \,\tilde{Y}^{-1} \, \widetilde{I}$ as  $\bar{I}_L^{\;T}\, ( \,\tilde{Y}_{GG}\,)^{-1} \,\bar{I}_L$, where $\tilde{Y}_{GG}$ is the Schur complement of $\tilde{Y}$}. 
\end{document}